# IR Spectroscopy of Synthetic Glasses with Mercury Surface Composition: Analogs for Remote Sensing

Corresponding Author: **Andreas Morlok**, Institut für Planetologie, Wilhelm-Klemm-Str. 10, 48149 Münster, Germany. Email: morlokan@uni-muenster.de, Tel. +49-251-83-39069

**Stephan Klemme**, Institut für Mineralogie, Corrensstraße 24, 48149 Münster, Germany. Email: stephan.klemme@uni-muenster.de

**Iris Weber**, Institut für Planetologie, Wilhelm-Klemm-Str. 10, 48149 Münster, Germany. Email: sonderm@uni-muenster.de

**Aleksandra Stojic**, Institut für Planetologie, Wilhelm-Klemm-Str. 10, 48149 Münster, Germany. Email: a.stojic@uni-muenster.de;

**Martin Sohn**, Hochschule Emden/Leer, Constantiaplatz 4, 26723 Emden, Germany, Email: martin.sohn@hs-emden-leer.de

**Harald Hiesinger**, Institut für Planetologie, Wilhelm-Klemm-Str. 10, 48149 Münster, Germany. Email: hiesinger@uni-muenster.de

**Abstract**

In a study to provide ground-truth data for mid-infrared observations of the surface of Mercury with the MERTIS (Mercury Radiometer and Thermal Infrared Spectrometer) instrument onboard the ESA/JAXA BepiColombo mission, we have studied 17 synthetic glasses. These samples have the chemical compositions of characteristic Hermean surface areas based on MESSENGER data.

The samples have been characterized using optical microscopy, EMPA and Raman spectroscopy. Mid-infrared spectra have been obtained from polished thin sections using Micro-FTIR, and of powdered size fractions of bulk material (0-25, 25-63, 93-125 and 125-250 µm) in the 2.5-18 µm range.

The synthetic glasses display mostly spectra typical for amorphous materials with a dominating, single Reststrahlen Band (RB) at 9.5 µm - 10.7 µm. RB Features of crystalline forsterite are found in some cases at 9.5-10.2 µm, 10.4-11.2 µm, and at 11.9 µm. Dendritic crystallization starts at a MgO content higher than 23 wt.% MgO.

The Reststrahlen Bands, Christiansen Features (CF), and Transparency Features (TF) shift depending on the $SiO_2$ and MgO contents. Also a shift of the Christiansen Feature of the glasses compared with the SCFM ($SiO_2/(SiO_2+CaO+FeO+MgO)$) index is observed. This shift could potentially help distinguish crystalline and amorphous material in remote sensing data. A comparison between the degree of polymerization of the glass and the width of the characteristic strong silicate feature shows a weak positive correlation.

A comparison with a high-quality mid-IR spectrum of Mercury shows some moderate similarity to the results of this study, but does not explain all features.

## 1. Introduction

Infrared spectroscopy allows determining the mineralogical composition of planetary surfaces via remote sensing. The Mercury Radiometer and Thermal Infrared Spectrometer (MERTIS) spectrometer onboard the future ESA/JAXA BepiColombo mission to Mercury will allow such remote sensing observations by mapping spectral features of the Hermean surface in the 7-14 μm range, with a spatial resolution of ~500 m (Benkhoff et al., 2010; Hiesinger et al., 2010). In order to correctly interpret remote sensing data, laboratory spectra of suitable analog material are of vital importance (Helbert et al., 2007; Maturilli et al., 2008). The IRIS (InfraRed and Raman for Interplanetary Spectroscopy) laboratory in Münster therefore generates spectra from analog material similar to those materials expected to occur on the surface of Mercury.

Surface regolith and exposed rocks of terrestrial planets and their moons are modified by impact events throughout their lifetimes (Hörz and Cintala, 1997). The investigation of how these related processes affect the spectral properties of the rocks is important for the correct interpretation of infrared data from planetary bodies. Higher impact shock, for example, results in amorphous phases produced in solid state transformation (such as maskelynite), or melt glass (e.g., Stöffler, 1966; Wünnemann et al., 2008; Osinski and Pierrazo, 2012; Jaret et al., 2015a). Under shock metamorphic conditions, minerals transform from crystalline to a solid amorphous state including diaplectic glasses like maskelynite at pressures of ~25 - ~40 GPa. Melting of feldspar starts at ~35 to ~45 GPa. Over 60 GPa rocks melt completely, which may result in quenched melt glass (e.g., Stöffler, 1966, 1971, 1984; Chao, 1967; von Engelhardt and Stöffler, 1968; Stöffler and Langenhorst, 1994; French, 1998; Johnson, 2012). Impact glass lacks a far-range order of its atomic constituents and represents the amorphous building block of a material, typically generated in events involving high shock pressure and temperatures (French, 1998; Speck et al., 2011).

In our study, we present the first mid-infrared reflectance data for synthetic glasses as analogs for melt glass based on the respective chemical compositions derived from remote sensing and model data for the surface of Mercury. Using synthetic materials allows us to produce more realistic analogs for Mercury surface rocks. To date, there are no Hermean meteorites we know of (Weber et al., 2016; Goodrich et al., 2017), chemical remote sensing data based on X-ray Spectrometer (XRS) and the Gamma-Ray and Neutron Spectrometer (GNRS) is the best information source available so far to deduce the surface composition of Mercury (Weider et al., 2015; vander Kaaden et al., 2016). We expect the surface rocks and regolith not only to consist of glassy material, a mixture of components of various shock stages seems more likely. Every respective shock stage will have different spectral characteristics (e.g. Morlok et al., 2016a, 2016b, 2017). Therefore, the spectra of the synthetic glasses produced in this study will serve as the endmember for studies of glass mixtures where amorphous and crystalline components are mixed to varying degrees. Also, areas that underwent more recent volcanism could be less affected by impact alteration. Such areas comprise large areas of the Mercurian surface (e.g. They could provide crystalline minerals (e.g. Deneva et al., 2013; Goudge et al., 2014; vander Kaaden et al., 2016).

To produce the synthetic glasses, we use the average chemical composition of surface regions identified in the MESSENGER data Compositions G1 and G2 (Charlier et al., 2013), which are the average compositions for larger areas in the equatorial region and the southern hemisphere of Mercury. They are distinguished by their variation in the Ca and Al contents (Tab.1) and cover both high-reflectance volcanic plains and low-reflectance rocks. Stockstill-Cahill et al. (2013) and Weider et al. (2012) present average compositions for the Mg poor, alkali-rich northern volcanic plains (NVPa) area, and the Mg-rich intercrater plains and heavily cratered terrain (IcP-HCTa) (Tab.1). Peplowski et al. (2015) present compositions of the high-Al and low-Mg Interior plains (CBC), i.e., the area inside the young Caloris impact crater and the low-Al and Mg-Northern Terrane (NC), i.e., northernmost part of Mercury above 60° northern latitude. Further areas are High-Mg Terranes (HMC) and an Intermediate composition (IC) (Tab.1).

Compositional data presented in vander Kaaden et al. (2015) and Weider et al. (2015) are the high-Mg (and low Al) region (HMR) and a sub region of the HMR with Ca and S enrichments (HMR-CaS), the low-Mg plains of the Caloris basin (CB), Mg-rich and poor parts of the northern volcanic plains (NP-HMg; NP-LMg, respectively), the Rachmaninoff basin (RaB), an area with high-Mg near the high-Al northern plains (HAl), a large pyroclastic deposit (PD), and the average of inter crater plains and various, cratered terrains (IT) (Weider et al., 2015).

We analyze four size fractions (0-25, 25-63, 63-125, 125-250 µm), motivated to better account for the high porosity and large grain size variations of surface regolith. Variation in grain size causes changes in the intensity of the characteristic Reststrahlen Bands (RB), fundamental mode absorption features in the 7-14 µm region, resulting in a loss of spectral contrast with decreasing grain size. An earlier study in the visible and near-infrared range (Sprague et al., 2007) indicated a high abundance of grains smaller 30 µm in size comprising the surface regolith on Mercury. Therefore, the corresponding RBs are expected to be weak in the remote sensing data of Mercury (Salisbury and Eastes, 1985; Salisbury and Wald, 1992; Mustard and Hayes, 1997). In addition, the transparency feature (TF), a characteristic additional spectral feature for small grain sizes, appears around 11-13 µm in the smallest grain size fractions below 50 µm (e.g., Salisbury, 1993). Potential TF features have been observed in ground based infrared observation of Mercury, indicating a high abundance of such fine-grained material in the regolith. This motivates the need for spectral data especially of the fine-grained size fractions (e.g., Cooper et al., 2001; Sprague et al., 2007).

Earlier reflectance and emission studies in the mid-infrared of synthetic glasses as analogues for impact melt glass were made by Byrnes et al. (2007) and Lee et al. (2010). They analyzed synthetic quartzofeldspathic glasses and found correlations between the band positions of characteristic dominant features and $SiO_2$ contents or Si/O ratios. Comparable results for synthetic glass with basaltic (low $SiO_2$) to intermediate (high $SiO_2$) composition were obtained by DuFresne et al. (2009), Minitti et al. (2002), and Minitti and Hamilton (2010). McMillan and Piriou (1982), Speck et al. (2011) and King et al (2004) provide additional overview of the infrared properties of silicate glass. Glasses from laser pulse experiments with a Martian soil analog JSC Mars-1 were analyzed by Basilevsky et al. (2000), Moroz et al. (2009), and Morris et al. (2000), resulting in spectra dominated by a strong single band in the 9.2-10.5 µm wavelength range. Earlier reflectance and emission studies of natural impact melt glass formed during impacts in the mid-infrared were performed by Thomson and Schultz (2002), Gucsik et al. (2004), Faulques et al. (2001), Fröhlich et al. (2013) and Morlok et al. (2016a and b). Spectra of these samples are dominated by a broad RB in the 8.9-10.3 µm range, with only few other features in the mid-infrared. A complementary study of silicate glasses with Mercurian and other planetary compositions was made by Cannon et al. (2016).

## 2. Samples and Techniques

### 2.1 Sample Compositions and Preparation of Glasses

The respective chemical composition used for the synthetic glass analogs of surface areas on Mercury are based on Charlier et al., 2013 (Char), Stockstill-Cahill et al., 2013 (Stock), Peplowski et al., 2015 (Pep), and vander Kaaden et al., 2015 (VdK) (comparable to those in vander Kaaden et al., 2016). The various studies and models did not always present the same range of oxide components, we therefore limited the composition used in this study to $SiO_2$, $TiO_2$, $Al_2O_3$, $Fe_2O_3$, MgO, $CaCO_3$, $Na_2O$, and $K_2O$ for better comparability. Components below 0.5 wt.% were omitted for individual mixtures for simplification. Starting material compositions are given in Table 1.

The oxide and carbonate starting mixtures were finely ground to a powder in an agate mortar under acetone and then dried. The resulting mixtures were placed in medium sized Pt crucibles in which they were slowly heated to 1000°C to de-carbonate. Subsequently, the mixtures were heated and melted in a conventional box furnace at 1450°C for 2h. They were quenched immediately after complete liquefaction, the crucibles were swiftly taken out of the furnace and submerged in water.  The samples were vitrified within 10 secs.

The samples were melted in a box furnace in air. Oxygen fugacity was not controlled. This may affect phase equilibria slightly, but only when high amounts of Fe, the only redox sensitive major element, are present.

Our samples were melted at high temperatures and kept at temperature for several hours. The melts are characterized by relatively low viscosity which ensures complete homogenization. Once melted, the structure of the melt does not depend on the starting material.

In two cases (ICP-HCTa (Stock) and RaB (VdK)), we re-heated the starting material at 1500°C in an attempt to remove crystals, which have formed during quenching in the first procedure.

In order to prepare grain size fractions, bulk glass material was ground in steel and agate mortars. The powder was cleaned in acetone and dry sieved for one hour to generate four size fractions: 0-25 µm, 25-63 µm, 63-125 µm, and 125-250 µm, by using an automatic Retsch Tap Sieve. In order to remove clinging fines, the larger two fractions were again cleaned in acetone. In addition, polished thick sections were prepared for microscopic investigation from the pure glass sample.

## 2.2. Optical Microscopy

Polarized light microscopy provides fast information about the crystalline or amorphous character of the single components. It also enables first mineral identification in the samples (Fig. 1), which is important for the subsequent Raman investigation to avoid mixed measurements on an inhomogeneous sample location. The first overview images of all polished thick sections were obtained with a KEYENCE Digital Microscope VHX-500F under normal light conditions and under crossed polarizers.

## 2.3 Raman Spectroscopy

All Raman measurements were conducted using an Ocean Optics IDR-Micro Raman system (IfP, Münster), operating with an OneFocus optical system equipped with a 40 x objective. The laser excitation is 532 nm and the spectral resolution is about 7 cm$^{-1}$. Spectra were obtained with a laser power of 1.8 mW starting at wavenumbers around 200 $cm^{-1}$. The spot size on the sample is approximately 2 µm in diameter. Every spectrum is the result of one measurement at 15 seconds acquisition time (Fig. 2a,b).

All spectra are automatically background and baseline subtracted and have not been smoothed. As usual, all spectra are given with arbitrary units, because the height of a signal only corresponds to the quality of the individual Raman scatterer itself (e.g. the double peak in olivine appears because the two $SiO_4$-stretching (v1 and v3) modes are active.) In addition, the Raman spectra of the glass are affected by fluorescence, which adds intensity in the form of an underlying continuum.

**2.4 Electron Microprobe Analysis**

Backscattered electron (BSE) images (Fig. 4a-c) show crystalline phases (brighter phases) formed during quenching. Detailed quantitative analyses of the glass and the olivines crystallized during the quenching process were made with a JEOL JXA-8530F Hyperprobe electron probe micro analyzer (EPMA) equipped with five wavelength dispersive spectrometers (WDS) (Fig.3). For the glass analyses, the probe was operated at an excitation voltage of 15 kV and a beam current of 5 nA. The beam diameter was defocused to 5 µm. The counting time was 5 seconds on the peak and 2 seconds on the background of each element, respectively. For mineral analyses we used an excitation voltage of 15 kV and a beam current of 15 nA with a slightly defocused beam diameter of 2 µm. The counting time for Mg, Al, Si, Ca, Fe, Ti, Cr, and Mn was 15 seconds on the peak and 5 seconds on the background. And, in order to avoid loss of the volatile elements Na and K, the counting time was reduced to 5 seconds on the peak and 2 seconds on the background for these two elements. The following natural and synthetic minerals with well-known compositions were used as standards: Jadeite ($Na_2O$), SanCarlos Olivine (MgO), Disthene ($Al_2O_3$), Hypersthene ($SiO_2$), Sanidine ($K_2O$), Diopside (CaO), Fayalite (FeO), Rutile ($TiO_2$), $Cr_2O_3$ ($Cr_2O_3$), and Rhodonite (MnO).

## 2.5 Bi-directional Diffuse Reflectance FTIR

For the bi-directional analyses of the sieved bulk powder size fractions we used aluminum sample cups with 1 cm diameter. The surface was gently flattened with a spatula following a procedure analog to that described by Mustard and Hayes (1997). For the bulk powder analyses in the mid-infrared from 2.5-18 μm, we used a Bruker Vertex 70 infrared system at IRIS laboratory.

We used a cooled MCT detector to ensure a high signal to noise ratio of the spectra. All analyses were made under low pressure ($10^{-3}$ bar). We accumulated 512 scans for each size fraction at al spectral resolution of 0.02 μm (compared to 0.2 μm for MERTIS; Hiesinger et al., 2010). The machine background was removed using a diffuse gold standard (INFRAGOLD[TM]). We obtained analyses with a variable geometry stage (Bruker A513) in order to emulate various observational geometries of an orbiter. The data presented in this study were obtained at 30° incidence (i) and 30° emergence angle (e).

The spectra returned from MERTIS will be emissivity data. For the comparison with remote sensing data in thermal infrared, reflectance and emission data have to be compared. This is usually done using Kirchhoff's law: ε = 1 − R (R=Reflectance, ε = Emission) (Nicodemus, 1965). For a direct comparison using Kirchhoff's law, the reflected light in all directions has to be collected. This relation works best for the comparison of directional emissivity and directional hemispherical reflectance. In our study, a bi-directional, variable mirror set-up was used, without a hemisphere integrating all reflected light. This has to be kept in mind when comparing the results in a quantitative manner with emission data (Salisbury et al., 1991; Hapke, 1993; Thomson and Salisbury, 1993; Salisbury et al., 1994; Christensen et al., 2001).

The spectral range of the MERTIS spectrometer is from 7-14 μm. Features at shorter and longer wavelengths can be of interest for other studies, we therefore present powder spectra from 6-18 μm (Fig. 5). The spectra are presented in reflectance, i.e., 0-1.

The width of infrared bands was determined using the FWHM (Full Width at Half Maximum) of the dominant spectral features. To derive this parameter from the often asymmetric, non-Gaussian bands, we used the Origin software. For the fitting, GCAS (Gram-Charlier peak function) and CCE (Chesler-Cram) fitting functions were applied to the spectra of the 125-250 μm size fraction, which were normalized to the same intensity.

The spectra presented in this study are accessible via an online database at the Institut für Planetologie in Münster (http://www.uni-muenster.de/Planetology/en/ifp/ausstattung/iris_spectra_database.html), and the Berlin Emissivity Database (BED).

## 2.6 In Situ FTIR Microscope

For in situ analyses, we used a Bruker Hyperion 2000 IR microscope attached to the external port of a Bruker Vertex 70v at the Hochschule Emden/Leer. Here we used a 256×256 μm² sized aperture to obtain analyses of small features with in situ reflectance spectroscopy on polished thin sections. For each spectrum, 128 scans were integrated. A gold mirror was used for background calibration (Fig. 6a,b).

## 3. Results

## 3.1 Optical Microscopy

The polished thick sections show a great diversity in color from nearly transparent samples, such as NC (Pep) (Fig. 1), to a brownish glass such as PD (VdK) as the darkest endmember. Samples exhibiting crystallites tend to be more brownish.

Several samples show heterogeneity in the form of clearly identifiable crystals, which are embedded in a still glassy matrix. These occur in the RaB (VdK) sample, while HMR (VdK), HMC (Pep), and HMR-CaS (VdK) are examples displaying high crystallinity (Fig. 1). In these samples, olivine crystals were identified by their habit and by their color in polarized light (Fig. 1).

## 3.2 Raman

Raman spectra of the pure glasses show typical shifts, two main broad signals, which cannot be attributed to single peaks, between 400 cm$^{-1}$ - 700 cm$^{-1}$ indicative of a silicate framework and between 850 cm$^{-1}$ - 1250 cm$^{-1}$, which is indicative of tetrahedrally coordinated cations (Fig. 2a; DiGenova et al., 2015). Raman shifts caused by glass are characterized by broad features lacking a crystal structure. The spectra in Fig. 2a are displayed with lower Mg contents on top and increasingly higher Mg contents. G2 (Char) is an outlier within this sequence, which might be a hint for a transitional sample showing incipient crystallization. Detailed Raman spectroscopy on the more heterogeneous samples, which include crystals confirms the existence of forsteritic olivine with a typical Raman double peak (DB) at 823 cm$^{-1}$ and 856 cm$^{-1}$ (Fig. 2b, blue line; Chopelas, 1991). In addition, the shift at around 700 cm$^{-1}$ (Fig.2b, green line) in the spectra of HMC (Pep) can be attributed to spinel (Downs, 2006).

## 3.3 EMPA

Results of the detailed quantitative analyses of the glasses as well as the sample weight of the starting oxides are given in Table 1. The analyses are average values of 20 – 70 measurements on each glass depending on the amount of olivine crystals. Analyses of the crystallized olivines, including their Fo-content, are listed in Table 2. The BSE image in Fig. 4a-c shows the dendritic growth of olivine as a result of quenching. In addition, as can be seen in Tables 1 and 2, the formation of olivine during quenching decreases with the decreasing MgO content in the original sample weight. The lower the MgO content the more identical are the quantitative analyses with the original values (Fig.3). At a MgO content higher than 23 wt.% (Tab.1), more and more olivine crystals are visible in the sample (Fig. 1). Also, chemical differences between the original sample (oxide mixture) and the glass increase above this threshold (Fig.3). However, a few crystallites are already visible at lower MgO contents like in Hal (VdK) (Fig.2a).

Normative calculations by vander Kaaden et al.(2016) also show high olivine contents for these samples.

Furthermore, BSE images combined with WDS analyses give hints on critical chemical thresholds for the formation of further crystalline phases. As visible in Figure 4c and 2b, spinel is another phase inside of an olivine grain with remnants of MgO around the spinel rim. Spinel and remnants of MgO occur in G1 (Char), HMR (VdK), HMR-CaS (VdK), and HMC (Pep). The threshold value for Mg for the crystallization of spinel and MgO remnants are less than 9 wt.% Al and more than 25 wt.% Mg(Tab.1).

**3.4 Diffuse Reflectance FTIR**

Spectra of the different size fractions are presented in the 6-18 μm range in Fig. 3, arranged according to increasing Mg-contents in the starting materials (Fig. 5, 6, Tab. 3). Features below 6 μm (2.9-3.5 μm) are either features of adsorbed volatiles on the starting materials or in the furnace and not shown.

For better comparison, the spectra of the largest size fraction (125-250 μm) are presented together in Fig. 6a.

A series of 12 samples with low Mg contents show spectra which are dominated by a single RB between 9.5 μm (NP LMg VdK) and 10.7 μm (ICP-HCTa Stock). Endmembers for the CF are 7.9 μm (NC Pep) and 8.3 μm (PD VdK), for the TF NP-LMg (12.4 μm VdK) with 11.8 μm and RaB (VdK) with 12.2 μm (Fig. 5). One exception is sample G2 (Char), while the sample has the single RB characteristic for this group, the feature (10.2-10.6 µm) is clearly shifted towards longer wavelength in comparison (Fig.6a).

The other exception among the low-Mg samples is NVPa (Stock), which shows a shoulder at 9.5-9.7 µm and the main band at 10.5-10.7 µm (Tab.3). This spectrum was grouped with the remaining four high-Mg samples HMR, HMR-CaS (VdK), HMC (Pep), and G1 (Char) that exhibit broader main RB features and show clear crystalline features at 9.5-9.9 µm, 10.1 µm, 10.5 - 11.9 µm (Fig.5). These are all typical forsterite bands (Hamilton, 2010).

The CF of these high-Mg samples ranges from 7.9 µm to 8.4 µm, and the TFs are found between 11.7 and 12.3 µm. Additional weak bands are located between 16.2 and 16.7 µm (Tab.3). Of the original experimental run, RaB (VdK) and ICP-HCTa (Stock) also exhibited crystalline features, which disappeared after repeated melting/quenching at temperatures over 1700°C.

**3.5 Micro-FTIR**

The FTIR spectra of the 'pure' glass samples are basically identical to those from the equivalent powder analyses of the same samples (Fig. 6a; Tab. 4). We also attempted to obtain spectra of glassy spots in samples showing crystallization (Fig. 6b; Tab. 4).

The CFs of pure glass and glass-spots range from 7.9 µm (NP-LMG; CBC and NC Pep) to 8.3 µm (G1, Char), the main feature is between 9.6 µm (NP-LMG VdK) to 10.1 µm (HMR-CaS) (Tab.4).

Spots from individual crystals and high-Mg samples show more variation. Features with varying intensity at ~9.6 µm, ~10.2 µm, 10.5-11.0µm, and 11.9 µm are typical olivine features (Hamilton, 2010; Lane et al., 2011), beginning with the HMR (VdK) sample (Tab.4).

**4. Discussion**

**4.1. Correlation of spectral and chemical features**

The aim of this study is to correlate spectral features of the laboratory analyses to remote sensing data obtained from the surface of Mercury. The spectra from observations of Mercury available often show low spectral contrast (e.g. Sprague et al., 1998). Therefore, it is useful to identify potential characteristic features that allow us to derive mineralogical and chemical information from remote sensing data even if the observational data is of comparatively low quality.

The CF, as a reflectance minimum, is a feature that can be easily identified in remote sensing data. Hence, the band position of the CF is commonly used as a proxy for the bulk chemical composition of samples, like the $SiO_2$ content (Salisbury 1993). Analyses of bulk powders and micro-FTIR are basically identical, and are consistent with the trend line for earlier studies (Morlok et al., 2016b) (Fig. 7). The analyses from an earlier study on natural impact glass (Morlok et al., 2016b) show a much wider variation from the trend line. This points towards a high purity of the synthetic material used in the current study, in contrast to the natural materials, which often contained fragments of unshocked or unmelted material (Morlok et al., 2016a and b).

If a strong RB is observable in remote sensing data, it also allows for characterizing the material (Fig. 8). Again, our band positions of the strong main RB features with respect to the $SiO_2$ contents fall close to observations in earlier studies (Lee et al., 2010). They demonstrate a better agreement than the natural impact glasses from our earlier study (Morlok et al., 2016b). At lower $SiO_2$ contents (below 60 wt.%) the results of this study show a similar trend than those for basaltic glass from DuFresne et al. (2009), where the trend line becomes much shallower. However, we cannot distinguish amorphous from crystalline material with this type of comparison.

The SCFM index ($SiO_2/(SiO_2+CaO+FeO+MgO)$) is a way to determine the degree of polymerization of a material, based on the increased interconnection of the $SiO_4$ tetrahedra (Walter and Salisbury, 1989). Earlier studies (Morlok et al., 2016b) have indicated that amorphous material like impact glasses plot off the trend line for crystalline terrestrial material (Cooper et al., 2002). The synthetic glasses in this study show a similar and in fact clearer behavior, continuing the trend line for basaltic compositions (Fig. 9).

A further way to correlate polymerization with spectral features is to use the bandwidth of the dominating silicate feature in the glass with the ratio between network forming cations (Si and Al) and network modifying anions (Fe, Ca, Mg, Na, K) (King et al., 2004; Dufresne et al., 2009; Speck et al., 2011). A comparison of this ratio to the FWHM (Full Width at Half Maximum) of only the glassy samples could help to obtain information about the degree of polymerization from remote sensing data. However, the correlation between these two parameters of the glass in our study is low with a correlation factor R (Pearson) of only 0.34 (Fig.10).

## 4.2. Comparison with Astronomical Observations of Mercury

Earth based mid-infrared observations of Mercury integrate large surface areas of up to $10^6$ $km^2$ and indicate a surface of mainly plagioclase with minor pyroxene (Donaldson-Hanna et al. 2007; Sprague et al., 1994, 2000, 2002, 2007; Sprague and Roush, 1998; Emery et al., 1998,; Cooper et al., 2001).

We compare our results with an astronomical spectrum (Fig. 11), obtained by the Mid-Infrared Array Camera (MIRAC) at the Kitt Peak Observatory. The spectrum is chosen due to its high signal to noise ratio and strong features in the spectral range of MERTIS (7-14 µm). The observed spectrum is the average of several observations of an area centered on ~210-250° longitude (Sprague et al., 2000). The baselined spectrum, re-calculated from emissivity to reflectance, shows a CF-like feature at 8.5 µm, strong RBs at 9.3 µm, 9.9 µm, and 11.0 µm, and a potential TF at 12.4 µm. Similarly, potential TFs in the 12.0-12.7 µm region are also observed in various other spectra of Mercury (Cooper et al., 2001; Sprague et al., 2007).

A comparison with results from this study demonstrates that glass from the 125-250 µm fraction of High Aluminium regions (HAl; VdK) and the 0-25 µm fraction of the High-Magnesium region HMC (VdK) (which already shows abundant crystallites) are the best to be compared to (Fig. 11).

The glassy HAl (VdK) analog sample has a main RB at the same position as in the astronomical spectrum (9.9 µm), although it is much broader. The HMC (VdK) analog sample has a TF at 12.3 µm, close to the band in the remote sensing spectrum (12.4 μm). Also, the CF is relatively close to the astronomical data, which show the CF at 8.4 μm. The broad RB feature of the Hal (VdK) analog sample around 11.2 μm is rather close to the (much narrower) 11.0 μm band in the Mercury spectrum. No equivalent for the 9.3 μm feature in the astronomical spectrum was found in the glasses of this study. This feature could indicate a pyroxene component (e.g., Hamilton, 2000; Sprague et al., 2000) in the surface regolith of Mercury.

However, we compare here laboratory analyses of samples with very distinct compositions to those of large surface areas of Mercury, effectively a 'bulk' spectra averaging many different mineral species. Future remote sensing data of smaller, chemically and mineralogically homogeneous areas of the surface of Mercury returned from the MERTIS instrument on BepiColombo (Benkhoff et al., 2010; Hiesinger et al., 2010) will allow to use the laboratory spectra much more efficiently.

Glass will probably mostly be contained in impactites and regolith, where it can be expected to be part of a mixture with other, crystalline materials (Tompkins and Peters, 2010; Morlok et al., 2016a and b). The large volcanic structures like smooth plains and pyroclastic deposits could be a source for crystalline material, which will make the detection of glasses difficult (Denevi et al., 2013; Goudge et al., 2014; vander Kaaden et al., 2016).

Visible and Near-Infrared studies of planetary glasses by Cannon et al. (2016) have shown that identifying glass in spectral mixtures is challenging even at mass abundances up to 70%. Especially mafic glasses are difficult to be identified 'by eye', but can be separated in spectral deconvolution. Studies in the mid-infrared range by Ramsey and Christensen (1998) were able to reproduce mixtures of crystalline minerals even for components below 10%. Still, this could lead to challenges identifying the glass in the surface regolith of Mercury. Surface regolith studies show up to 45% agglutinitic glass, much higher than 2-5% observed for glass on the lunar surface (Warrell et al., 2010).

Also, space weathering has to be taken into account in this context, which can be expected much more intense closer to the sun and also produce glassy materials (Papike et al., 1982; Hapke, 2001, Sprague et al., 2007).

## 5. Summary and Conclusions

The study of a series of synthetic glasses based on surface regions of Mercury displays mostly spectra typical for amorphous materials, characterized by a single RB between 9.5 µm and 10.7 µm. Several spectra show features of crystalline olivine species in both the powder and micro-FTIR spectra.

FTIR features of crystalline species at 9.5-10.2 µm, 10.4-11.2 µm, and at 11.9 µm in the powder and micro-FTIR spectra are mostly characteristic of forsterite, appearing in a smaller number of samples at higher MgO contents, beginning with the HMR (VdK) sample. Optical, Raman and EMPA data have signs of crystallization starting with the RaB (VdK) material. This points towards a threshold for crystallization starting at MgO contents higher than 23wt.% MgO.

However, in Raman spectra as well as in FTIR G2 (Char) is an outlier regarding the position of the strongest feature. This might be indicative of a starting crystallization at a lower MgO concentration in this sample.

The RBs, CFs, and TFs shift, depending on the $SiO_2$ and MgO contents and exhibit similarity to earlier studies. We also confirm a shift of the CF position of amorphous material compared with the respective crystalline SCFM index, which could help distinguishing crystalline and amorphous material in remote sensing data in the mid-IR.

A comparison between the degree of polymerization of the glass and the width of the characteristic strong silicate feature shows only slight correlation. However, here sample heterogeneity – such as incipient crystallization – probably affects the spectral features.

A comparison with a high-quality mid-IR spectrum of Mercury shows some similarity to the results of this study, but does not explain all features. A series of analyses of distinct size fractions is needed to distinguish between genuine effects and effects induced by mixed grain size fractions in the natural regolith. However, since regolith is also an intimate mixture of many phases, spectral unmixing modelling will also have to take many other grain-size fractions of other potential mineral phases into account.

## 6. Acknowledgements

We thank Isabelle Dittmar (Emden) for analytical support, Ulla Heitmann (Münster) for thin section preparation. This work was partly supported by DLR grant 50 QW 0901 in the framework of the BepiColombo mission.

**References**

Basilevsky A.T., Yakovlev O.I., Fisenko A.V., Semjonova L.F., Moroz L.V., Pieters C.M., Hiroi T., Zinovieva N.G., Keller H.U., Semenova A.S., Barsukova L.D., Roshchina I.A.,  Galuzinskaya A.K. Stroganov I.A. (2000) Simulation of Impact Melting Effect on optical Properties of Martian Regolith. 31st Annual Lunar and Planetary Science Conference, Abstract no. 1214.

Benkhoff J., van Casteren J., Hayakawa H. Fujimoto M., Laakso H., Novara M., Ferri P., Middleton H. R., Ziethe R. (2010) BepiColombo---Comprehensive exploration of Mercury: Mission overview and science goals. Planetary and Space Science 58, 2-20.

Byrnes J.M., Ramsey M.S., King P.L., Lee L.J. (2007) Thermal infrared reflectance and emission spectroscopy of quartzofeldspathic glasses. Geophysical Research Letters 34, L01306.

Cannon, K.M., Mustard J.F., Parman S.W., Sklute E.C., Dyar M.D., Cooper R.F. (2017) Spectral properties of Martian and other planetary glasses and their detection in remotely sensed data. Journal of Geophysical Research: Planets 122, 249-268.

Chao E. C. T. (1967) Impact metamorphism. In Researches in Geochemistry, Vol. 2 (P. H. Abelson, ed.), pp. 204–233. John Wiley and Sons, New York.

Charlier B., Grove T. L., Zuber M. T. (2013) Phase equilibria of ultramafic compositions on Mercury and the origin of the compositional dichotomy. Earth and Planetary Science Letters 363, 50-60.

Chopelas A. (1991) Single crystal Raman spectra of forsterite, fayalite, and monticellite. American Mineralogist 76, 1101 – 1109.

Christensen P.R., Bandlield J.L, Hamilton V.E. (2001) Mars Global Surveyor Thermal Emission Spectrometer experiment: Investigation description and surface science results. Journal of Geophysical Research 106 (ElO), 23823-23872.

Cooper B., Potter A., Killen R., Morgan T. (2001) Midinfrared spectra of Mercury. Journal of Geophysical Research 106, E12, 32803-32814.

Cooper B.L., Salisbury J.W., Killen R.M., Potter, A.E. (2002) Midinfrared spectral features of rocks and their powders. Journal of Geophysical Research 107, E4, 5017-5034.

DiGenova D., Morgavi D., Hess K.-W., Neuville D.R., Borovkov N., Perugini D., and Dingwell D.B. (2015) Approximate chemical analysis of volcanic glasses using Raman spectroscopy. Journal of Raman spectroscopy 46(12), 1235 – 1244.

Denevi, B.W., Ernst C.M., Meyer H.M., Robinson M.S., Murchie S.L., Whitten J.L., Head J.W., Watters T.R., Solomon S.C., Ostrach L.R. et al. (2013) The distribution and origin of smooth plains on Mercury. Journal of Geophysical Research: Planets 118,. 891-907.

Donaldson-Hanna K.L, Sprague A.L., Kozlowski R.W.H., Boccafolo K., Warell J. (2007) Mercury and the Moon: Initial Findings from Mid-Infrared Spectroscopic Measurements of the Surface. 38th Lunar and Planetary Science Conference, Abstract 1338

Downs R.T. (2006) The RRUFF Project: an integrated study of the chemistry, crystallography, Raman and infrared spectroscopy of minerals. 19[th] General Meeting of the International Mineralogical Association, Kobe, Japan. O03-13.

DuFresne C.D.M., King P.L., Dyar D., Dalby K.N. (2009) Effect of $SiO_2$, total FeO, $Fe^{3+}/Fe^{2+}$, and alkali elements in basaltic glasses on mid-infrared spectra. American Mineralogist 94, 1580-1590.

von Engelhardt W., Stöffler D. (1968) Stages of shock metamorphism in crystalline rocks of the Ries Basin, Germany. In Shock Metamorphism of Natural Materials (B. M. French and N. M. Short, eds.), pp. 159–168. Mono Book Corp., Baltimore.

Emery  J.P., Sprague A.L., Witteborn F.C., Colwell F.C., Kozlowski R.W.H. (1998) Mercury: Thermal Modeling and Mid-infrared (5–12 μm) Observations. Icarus 136, 104-123.

Faulques E., Fritsch E., Ostroumov M. (2001) Spectroscopy of natural silica-rich glasses. Journal of Mineralogical and Petrological Sciences 96, 120-128.

French B.M. (1998) Traces of Catastrophe: A Handbook of Shock-Metamorphic Effects in Terrestrial Meteorite Impact Structures. LPI Contribution No. 954, Lunar and Planetary Institute, Houston. 120 pp

Fröhlich F., Poupeau G., Badou A., Le Bourdonnec F.X., Sacquin Y., Dubernet S., Bardintzeff J.M., Veran M., Smith D.C., Diemer E. (2013) Libyan Desert Glass: New field and Fourier transform infrared data. Meteoritics & Planetary Science 48, 2517-2530.

Goodrich C.A., Kita N.T., Yin Q.-Z., Sanborn M.E., Williams C.D., Nakashima D., Lane M.D., Boyle S. (2017) Petrogenesis and Provenance of Ungrouped Achondrite Northwest Africa 7325 from Petrology, Trace Elements, Oxygen, Chromium and Titanium Isotopes, and Mid-IR Spectroscopy. Geochimica et Cosmochimica Acta, Accepted Manuscript in Press.

Goudge, T.A., Head J.W., Kerber L., Blewett D.T., Denevi B.W., Domingue D.L., Gillis-Davis J.J., Gwinner K., Helbert J., Holsclaw G.M. et al. (2014) Global inventory and characterization of pyroclastic deposits on Mercury: New insights into pyroclastic activity from MESSENGER orbital data. Journal of Geophysical Research: Planets 119, 635-658.

Gucsik A., Koeberl C., Brandstätter F., Libowitzky E., Zhang M. (2004) Infrared, Raman, and cathodoluminescence studies of impact glasses. Meteoritics & Planetary Science 39, 1273-1285.

Hamilton V.E. (2000) Thermal infrared emission spectroscopy of the pyroxene mineral series. Journal of Geophysical Research 105 (E4), 9701-9716.

Hamilton V.E. (2010) Thermal infrared (vibrational} spectroscopy of Mg-Fe olivine a review and applications to determining the composition of planetary surfaces. Chemie der Erde - Geochemistry 70, 7-33.

Hapke B. (1993), Theory of Reflectance and Emittance Spectroscopy. Cambridge Univ. Press, New York.

Hapke B. (2001) Space weathering from Mercury to the asteroid belt. Journal of Geophysical Research 106, 10039-10073.

Helbert, J.; Moroz, L. V.; Maturilli, A.; Bischoff, A.; Warell, J.; Sprague, A.; Palomba, E. (2007) A set of laboratory analogue materials for the MERTIS instrument on the ESA BepiColombo mission to Mercury. Advances in Space Research 40, 272-279.

Hiesinger H., Helbert J., Mertis Co-I Team (2010) The Mercury Radiometer and Thermal Infrared Spectrometer (MERTIS) for the BepiColombo mission. Planetary and Space Science 58, 144–165.

Hörz F., Cintala M. (1997) Impact experiments related to the evolution of planetary regoliths. Meteoritics & Planetary Science 32, 179-209.

Jaret S.J., Woerner W.R., Philips B.L., Ehm L., Nekvasil H., Wright S.P., Glotch T.D. (2015a) Maskelynite formation via solid-state transformation: Evidence of infrared and X-ray anisotropy.  Journal of Geophysical Research: Planets 120, 570-587.

Johnson J.R. (2012) Thermal infrared spectra of experimentally shocked andesine anorthosite. Icarus 221, 359–364.

King P.L., McMillan P.F., Moore G.M. (2004) Infrared spectroscopy of silicate glasses with application to natural systems, in Infrared Spectroscopy in Geochemistry, Exploration Geochemistry, and Remote Sensing, Mineral. Assoc. of Can. Short Course Ser., vol. 33, edited by P. L. King, M. S. Ramsey, and G. A. Swayze, pp. 93– 133, Mineral. Assoc. of Can., Ottawa.

Lane M.D., Glotch T.D., Dyar M.D., Pieters C.M., Klima R., Hiroi T., Bishop J.L., Sunshine J. (2011) Midinfrared spectroscopy of synthetic olivines: Thermal emission, specular and diffuse reflectance, and attenuated total reflectance studies of forsterite to fayalite. Journal of Geophysical Research (Planets) 116, E8, CiteID E08010.

Lee R.J., King P.L., Ramsey M.S. (2010) Spectral analysis of synthetic quartzofeldspathic glasses using laboratory thermal infrared spectroscopy. Journal of Geophysical review 115, B06202.

Maturilli A., Helbert J., Moroz L. (2008) The Berlin emissivity database (BED). Planetary and Space Science 56, 420-425.

McMillan P. and Piriou B. (1982) The structures and vibrational spectra of crystals and glasses in the silica-alumina system. Journal of Non-Crystalline Solids 53, 279–298.

Minitti M.E., Mustard J.F., Rutherford M.J. (2002) Effects of glass content and oxidation the spectra of SNC-like basalts: Applications to Mars remote sensing. Journal of Geophysical Research (Planets) 107, 6-11.

Minitti M.E., Hamilton V.E. (2010) A search for basaltic-to-intermediate glasses on Mars: Assessing martian crustal mineralogy. Icarus 210, 135-149.

Morlok A., Stojic A., Dittmar I., Hiesinger H., Tiedeken M., Sohn M., Weber I., Helbert J. (2016a) Mid-infrared spectroscopy of impactites from the Nördlinger Ries impact crater. Icarus 264, 352-368.

Morlok A., Stojic A., Weber I., Hiesinger H., Zanetti M., Helbert J. (2016b) Mid-infrared bi-directional reflectance spectroscopy of impact melt glasses and tektites. Icarus 278, 162-179.

Morlok A., Bischoff A., Patzek M., Sohn M., Hiesinger H. (2017) Chelyabinsk - a rock with many different (stony) faces: An infrared study. Icarus 284, 431-442.

Moroz L.V., Basilevsky A.T., Hiroi T., Rout S.S., Baither D, van der Bogert C.H., Yakovlev O.I., Fisenko A.V., Semjonova L.F., Rusakov V.S., Khramov D.A., Zinovieva N.G., Arnold G., Pieters C.M. (2009) Spectral properties of simulated impact glasses produced from martian soil analogue JSC Mars-1. Icarus 202, 336-353

Morris R. V., Le L., Lane M. D., Golden D. C., Shelfer T. D., Lofgren G. E., Christensen P. R. (2000) Multidisciplinary Study of Synthetic Mars Global Average Soil Glass. 31st Annual Lunar and Planetary Science Conference, Abstract no. 1611

Mustard J.F., Hays J.E. (1997) Effects of Hyperfine Particles on Reflectance Spectra from 0.3 to 25 μm. Icarus 125, 145-163.

Nicodemus F. E. (1965) Directional reflectance and emissivity of an opaque surface. Applied Optics 4, 767-773.

Osinksi R.R., Pierazzo E. (2012) Impact cratering: processes and products. In: Impact Cratering: Processes and Products (Eds. G.R.Osiniski and E.Pierazzo), pp 1-17; Wiley-Blackwell.

Papike J.J., Simon S.B., White C., Laul J.C. (1982) The relationship of the lunar regolith less than 10 micrometer fraction and agglutinates. I - A model for agglutinate formation and some indirect supportive evidence. In: Lunar and Planetary Science Conference, 12th, Houston, TX, March 16-20, 1981, Proceedings,  New York and Oxford, Pergamon Press, 1982, p. 409-420.

Peplowski P.N., Lawrence D.J., Feldman W.C., Goldsten J.O., Bazell D., Evans L.G., Head J.W., Nittler L.R., Solomon S.C., Weider S.Z. (2015) Geochemical terranes of Mercury's northern hemisphere as revealed by MESSENGER neutron measurements. Icarus 253, 346-363.

Ramsey M.S., Christensen P.R. (1998) Mineral abundance determination: Quantitative deconvolution of thermal emission spectra. Journal of Geophysical Research 103, 577-596

Salisbury J.W., Eastes J.W. (1985) The Effect of Particle Size and Porosity on Spectral Contrast in the Mid-lnfrared. Icarus 64, 586-588

Salisbury J. W., Walter L. S., Vergo N., and D'Aria D. M. (1991) Infrared (2.1–25 lm) spectra of minerals. Baltimore, MD: Johns Hopkins University Press.

Salisbury J.W., Wald A. (1992) The Role of Volume Scattering in Reducing Spectral Contrast of Reststrahlen Bands in Spectra of Powdered Minerals. Icarus 96, 121-128.

Salisbury J.W. (1993) Mid-Infrared Spectroscopy: Laboratory Data. In: Remote Geochemical Analysis: Elemental and Mineralogical Composition, Eds: C.M. Pieters and P.A.J. Englert. Cambridge University Press.

Salisbury J.W., Wald A., Di Aria D.M. (1994) Thermal-infrared remote sensing and Kirchhoff's law 1. Labaratory measurements. Journal of Geophysical Research 99, B6, 11897-11911.

Speck A.K., Whittington A.G., Hofmeister A.M. (2011) Disordered Silicates in Space: A Study of Laboratory Spectra of "Amorphous" Silicates. The Astrophysical Journal 740, 1-17.

Sprague A. L., Kozlowski R. W. H., Witteborn F. C., Cruikshank D. P., and Wooden D. H. (1994). Mercury: Evidence for anorthosite and basalt from mid-infrared (7.5–13.5 micrometer) spectroscopy. Icarus 109, 156–167.

Sprague A. L. and Roush T. L. (1998). Comparison of laboratory emission spectra with Mercury telescopic data. Icarus 133,174–183.

Sprague A., Deutsch L. K., Hora J., Fazio G. G., Ludwig B., Emery J., and Hoffmann W. F. (2000). Mid-infrared (8.1– 12.5 μm) imaging of Mercury. Icarus 147, 421–432.

Sprague A.L., Emery J.P., Donaldson K.L., Russell R.W., Lynch D.K., Mazuk A.L. (2002) Mercury: Mid-infrared (3-13.5 μm) observations show heterogeneous composition, presence of intermediate and basic soil types, and pyroxene. Meteoritics&Planetary Science 37, 1255-1268.

Sprague A., Warell J.M., Cremonese G., Langevin Y., Helbert J., Wurz P., Veselovsky I., Orsini S., Milillo A. (2007)  Mercury's Surface Composition and Character as Measured by Ground-Based Observations. Space Science Reviews 132, 399-431

Stockstill-Cahill K.R., McCoy T.J., Nittler L.R., Weider S.Z., Hauck S.A.II (2013) Magnesium-rich crustal compositions on Mercury: Implications for magmatism from petrologic modeling. Journal of Geophysical Research, Volume 117, CiteID E00L15

Stöffler D. (1966) Zones of impact metamorphism in the crystalline rocks at the Nördlinger Ries Crater. Contributions to Mineralogy and Petrology, 12, 15–24.

Stöffler D. (1971) Progressive metamorphism and classification of shocked and brecciated crystalline rocks at impact craters. Journal of Geophysical Research 76, 5541–5551.

Stöffler D. (1984) Glasses formed by hypervelocity impact. Journal Non-Crystalline Solids 67, 465–502.

Stöffler D. and Langenhorst F. (1994) Shock metamorphism of quartz in nature and experiment: I. Basic observation and theory. Meteoritics 29, 155–181.

Thomson J. L., Salisbury J.W. (1993) The mid-infrared reflectance of mineral mixtures (7-14 microns). Remote Sensing of Environment 45, 1-13.

Thomson B.J., Schultz P.H. (2002) Mid-infrared spectra of Argentine impact melts: Implications for Mars. 33rd Annual Lunar and Planetary Science Conference, Abstract 1595.

Tompkins S., Pieters C.M. (2010) Spectral characteristics of lunar impact melts and inferred mineralogy. Meteoritics & Planetary Science 45, 1152-1169.

Vander Kaaden K.E., McCubbin F.M., Nittler L.R., Weider S.Z. (2015) Petrologic Diversity of Rocks on Mercury. 46th Lunar and Planetary Science Conference, Abstract #1364

Vander Kaaden K. E., McCubbin F. M., Nittler L. R., Peplowski P. N., Weider S. Z., Frank E. A., McCoy T. (2016) Geochemistry, Mineralogy, and Petrology of Boninitic and Komatiitic Rocks on the Mercurian Surface: Insights into the Mercurian Mantle. Icarus (in Press http://dx.doi.org/10.1016/j.icarus.2016.11.041)

Walter L.S., Salisbury J.W. (1989) Spectral characterization of igneous rocks in the 8- to 12-micron region. Journal of Geophysical Research 94, 9203-9213.

Weber I., Morlok A., Bischoff A., Hiesinger H., Ward D., Joy K. H., Crowther, S. A., Jastrzebski N. D., Gilmour J. D., Clay P. L. et al. (2016) Cosmochemical and spectroscopic properties of Northwest Africa 7325—A consortium study. Meteoritics & Planetary Science 51, 3-30.

Weider, Shoshana Z.; Nittler, Larry R.; Starr, Richard D.; McCoy, Timothy J.; Stockstill-Cahill, Karen R.; Byrne, Paul K.; Denevi, Brett W.; Head, James W.; Solomon, Sean C. (2012) Chemical heterogeneity on Mercury's surface revealed by the MESSENGER X-Ray Spectrometer. Journal of Geophysical Research (Planets) 117, E00L05

Weider, Shoshana Z.; Nittler, Larry R.; Starr, Richard D.; Crapster-Pregont, Ellen J.; Peplowski, Patrick N.; Denevi, Brett W.; Head, James W.; Byrne, Paul K.; Hauck, Steven A.; Ebel, Denton S.; Solomon, Sean C. (2015) Evidence for geochemical terranes on Mercury: Global mapping of major elements with MESSENGER's X-Ray Spectrometer. Earth and Planetary Science Letters 416, 109-120.

Wünnemann K., Collins G. S., Osinski G. R. (2008) Numerical modelling of impact melt production in porous rocks. Earth and Planetary Science Letters 269, 530-539.

**Figure Captions**

Figure 1: Overview of optical images of polished blocks of the samples, arranged according to increasing Mg contents. Increasing abundance of Mg is correlated with increasing appearance of crystallites in the quenched melt glass. The sample for NVPa (Stock) was only available in powdered form. Red squares are 256×256 μm sized areas analyzed with micro-FTIR. Red scale bars in the images for NVPa (Stock), ICP-HCTa (Stock), and RaB (VdK) represent 256 μm.

Figure 2a: Raman spectra for the glasses, in the range from 300 $cm^{-1}$ to 1300 $cm^{-1}$ where the Mg-abundances increase from the top to the bottom. The broad peak between 400 $cm^{-1}$ and 700$^{-1}$ is a result from the extant silicate framework and between 850 $cm^{-1}$ and 1250 $cm^{-1}$ tetrahedra coordinated cations are present. Intensity in graphs is normalized on the strongest signal. The spectra are shifted vertically for a better view.

Figure 2b: Raman spectra for the crystalline phases formed during quenching in the range from 400 $cm^{-1}$ to 1600 $cm^{-1}$. The blue line represents a typical olivine spectrum with the DB at 823 $cm^{-1}$ and 856 $cm^{-1}$. Beyond this the green line shows an additional peak at 700 $cm^{-1}$ fitting to spinel. Intensity in graphs is normalized on the strongest signal. The spectra are shifted vertically for a better view.

Figure 3: Changes in main element abundances ($SiO_2$, $Al_2O_3$, MgO, CaO, FeO) between starting material (Initial) and quenched glass (EMPA data). Data is ordered by their increasing Mg content. Most samples with lower Mg abundance show only slight variation. Only high-Mg samples that produced crystallites show increasing differences.

Figure 4: Representative SEM BSE images of crystallites.

a) Example of the dendritic growth of olivine (Ol) as result of quenching in G1 (Char).

b) Example of the dendritic growth of olivine (Ol) as result of quenching in HAI (VdK).

c) Example of a spinel (light grey; Sp) as another present phase inside of an olivine grain (dark grey; Ol) with remnants of MgO (white) around the spinel rim in G1 (Char).

Figure 5: FTIR reflectance spectra of powdered glass in four size fractions (0-25 µm, 25-63 µm, 63-125 µm, 125-250 µm), sorted by increasing Mg-content. The crystalline features appearing at higher Mg-contents are forsterite bands (Hamilton, 2010).

Figure 6a: Comparison of FTIR reflectance powder spectra of the 125-250 µm size fraction of all samples. Mg-abundance is increasing from top to bottom. Vertical lines denote important features: Christiansen Feature (CF), and Reststrahlenband (RB). A shift of these features to longer wavelengths is correlated with increasing Mg abundance. Circle: olivine features in high-Mg samples (Hamilton, 2010).

Figure 6b: Micro-FTIR reflectance spectra of polished samples. Spectra are sorted by increasing Mg-content. Vertical lines denote important features: Christiansen Feature (CF) and Reststrahlenband (RB). A shift of these features to loner wavelengths is correlated with increasing Mg abundance. Left column: Spectra of glass or glassy regions in samples showing crystallization. Right column: Spectra of areas with high contents of crystallites. Vertical lines show typical forsterite RB bands.

Figure 7: $SiO_2$ contents of glasses (in wt.%) compared with the position of the Christiansen Feature (in µm). The samples of this study fall on the dashed trend line defined by earlier studies on impact glasses (Morlok et al., 2016) and crystalline terrestrial rocks (Cooper et al., 2002).

Figure 8: $SiO_2$ contents of glasses (in wt.%) compared with the position of the strongest Reststrahlen-Band (in µm). Glasses from this study are similar to earlier studies (Cooper et al., 2002), but also diverge from the trend lower $SiO_2$ contents similar to basaltic glasses in DuFresne et al., 2009.

Figure 9: Comparison of the SCFM index ($SiO_2$/($SiO_2$+CaO+FeO+MgO); Walter and Salisbury, 1989): with the position of the CF (in µm). The results of the glasses in this study confirm the findings in Morlok et al. (2016), where glassy material plotted below the trend line for crystalline materials. The trend line is for analyses of terrestrial rocks (Cooper et al., 2002).

Figure 10: Comparison of the Full Width at Half Maximum (FWHM) of the strong silicate feature in glasses with the degree in polymerization, based on a ratio between network building ions (Si, Al) and network modifiers (Fe, Ca, Mg, Na, K).

Figure 11: Comparison of a Mercury mid-infrared spectrum obtained by ground based observation (Sprague et al., 2000) with the two most similar spectra from this study, a high-Al region (HAl VdK) and the finest size fraction of a high Mg-region (HMC VdK). There is a general similarity for the CF and TF as well as the RBs at 9.9 μm, 11.2 μm. There is not equivalent for the feature at 9.3 μm.

| | NP-LMg (VdK) | | | NC (Pep) | | | CBC (Pep) | | |
|---|---|---|---|---|---|---|---|---|---|
| | Glass | Standard deviation | Start Material | Glass | Standard deviation | Start Material | Glass | Standard deviation | Start Material |
| $Na_2O$ | 4.65 | 0.14 | 4.65 | 7.46 | 0.21 | 7.42 | 3.65 | 0.21 | 3.72 |
| $MgO$ | 11.54 | 0.23 | 11.73 | 11.65 | 0.21 | 12.00 | 12.90 | 0.39 | 12.81 |
| $Al_2O_3$ | 11.12 | 0.21 | 11.03 | 11.60 | 0.19 | 11.48 | 17.27 | 0.24 | 17.20 |
| $SiO_2$ | 63.34 | 0.31 | 63.62 | 62.14 | 0.44 | 61.95 | 58.60 | 0.43 | 59.00 |
| $K_2O$ | 0.05 | 0.03 | | 0.05 | 0.04 | | 0.04 | 0.03 | |
| $CaO$ | 5.69 | 0.20 | 5.72 | 5.54 | 0.17 | 5.67 | 5.79 | 0.20 | 5.79 |
| $FeO$ | 1.81 | 0.16 | 1.76 | 0.76 | 0.13 | 0.75 | 0.81 | 0.12 | 0.70 |
| $TiO_2$ | 1.56 | 0.22 | 1.49 | 0.60 | 0.20 | 0.58 | 0.51 | 0.17 | 0.56 |
| $Cr_2O_3$ | 0.02 | 0.04 | | 0.02 | 0.03 | | 0.02 | 0.03 | |
| $MnO$ | 0.02 | 0.03 | | 0.02 | 0.03 | | 0.03 | 0.04 | 0.14 |
| Total | 99.81 | | 100.00 | 99.83 | | 100.00 | 99.62 | | 100.00 |

| | CB (VdK) | | | NVPa (Stock) | | | NP-HMg (VdK) | | |
|---|---|---|---|---|---|---|---|---|---|
| | Glass | Standard deviation | Start Material | Glass | Standard deviation | Start Material | Glass | Standard deviation | Start Material |
| $Na_2O$ | 4.02 | 0.16 | 4.13 | 0.06 | 0.03 | | 5.95 | 0.15 | 6.18 |
| $MgO$ | 13.69 | 0.32 | 13.17 | 16.09 | 0.28 | 16.51 | 17.48 | 0.36 | 17.65 |
| $Al_2O_3$ | 15.98 | 0.22 | 15.88 | 15.42 | 0.19 | 15.38 | 13.01 | 0.23 | 12.62 |
| $SiO_2$ | 58.13 | 0.39 | 59.35 | 58.17 | 0.32 | 58.82 | 57.01 | 0.41 | 57.00 |
| $K_2O$ | 0.04 | 0.04 | | 0.05 | 0.02 | | 0.06 | 0.03 | |
| $CaO$ | 5.96 | 0.26 | 6.01 | 4.91 | 0.14 | 4.84 | 5.02 | 0.22 | 5.22 |
| $FeO$ | 1.09 | 0.16 | 0.99 | 3.65 | 0.21 | 3.54 | 0.09 | 0.07 | |
| $TiO_2$ | 0.67 | 0.21 | 0.46 | 0.94 | 0.20 | 0.92 | 1.14 | 0.22 | 1.33 |
| $Cr_2O_3$ | 0.03 | 0.03 | | 0.01 | 0.01 | | 0.02 | 0.03 | |
| $MnO$ | 0.02 | 0.03 | | 0.01 | 0.01 | | 0.03 | 0.03 | |
| Total | 99.64 | | 100.00 | 99.31 | | 100 | 99.81 | | 100.00 |

| | G2 (Char) | | | HAI (VdK) | | | IC (Pep) | | |
|---|---|---|---|---|---|---|---|---|---|
| | Glass | Standard deviation | Start Material | Glass | Standard deviation | Start Material | Glass | Standard deviation | Start Material |
| $Na_2O$ | 0.04 | 0.04 | | 4.19 | 0.16 | 4.11 | 3.39 | 0.18 | 3.44 |
| $MgO$ | 18.78 | 0.30 | 19.3 | 18.73 | 0.53 | 19.42 | 19.96 | 0.40 | 19.85 |
| $Al_2O_3$ | 13.20 | 0.23 | 13.2 | 15.94 | 0.30 | 15.79 | 14.07 | 0.21 | 13.95 |
| $SiO_2$ | 56.15 | 0.39 | 56.6 | 53.06 | 0.52 | 53.00 | 53.87 | 0.64 | 54.46 |
| $K_2O$ | 0.04 | 0.03 | | 0.04 | 0.03 | | 0.04 | 0.03 | |
| $CaO$ | 7.10 | 0.22 | 6.92 | 6.52 | 0.21 | 6.43 | 5.86 | 0.18 | 5.69 |
| $FeO$ | 3.49 | 0.13 | 3.37 | 0.09 | 0.07 | | 2.12 | 0.17 | 1.97 |
| $TiO_2$ | 0.68 | 0.17 | 0.59 | 1.29 | 0.32 | 1.24 | 0.54 | 0.20 | 0.51 |
| $Cr_2O_3$ | 0.02 | 0.04 | | 0.02 | 0.03 | | 0.02 | 0.03 | |
| $MnO$ | 0.05 | 0.04 | | 0.02 | 0.03 | | 0.03 | 0.04 | 0.13 |
| Total | 99.52 | | 99.98 | 99.91 | | 100.00 | 99.90 | | 100.00 |

IT (VdK)  PD (VdK)  ICP-HCTa (Stock)

|  | Glass | Standard deviation | Start Material | Glass | Standard deviation | Start Material | Glass | Standard deviation | Start Material |
|---|---|---|---|---|---|---|---|---|---|
| Na$_2$O | 3.55 | 0.16 | 3.55 | 4.04 | 0.21 | 4.06 | 0.05 | 0.03 | |
| MgO | 21.01 | 0.30 | 21.30 | 21.21 | 0.24 | 21.51 | 22.50 | 0.26 | 22.86 |
| Al$_2$O$_3$ | 14.02 | 0.24 | 13.99 | 11.67 | 0.22 | 11.55 | 12.80 | 0.31 | 12.62 |
| SiO$_2$ | 52.14 | 0.33 | 52.62 | 51.91 | 0.41 | 52.34 | 53.32 | 0.44 | 54.21 |
| K$_2$O | 0.04 | 0.03 | | 0.05 | 0.04 | | 0.04 | 0.03 | |
| CaO | 5.94 | 0.22 | 5.74 | 7.47 | 0.21 | 7.36 | 6.28 | 0.21 | 6.20 |
| FeO | 1.83 | 0.15 | 1.58 | 2.14 | 0.18 | 1.95 | 3.64 | 0.21 | 3.26 |
| TiO$_2$ | 1.22 | 0.25 | 1.23 | 1.23 | 0.26 | 1.22 | 0.94 | 0.23 | 0.85 |
| Cr$_2$O$_3$ | 0.03 | 0.04 | | 0.03 | 0.03 | | 0.05 | 0.04 | |
| MnO | 0.02 | 0.03 | | 0.03 | 0.04 | | 0.02 | 0.02 | |
| Total | 99.80 | | 100.00 | 99.77 | | 100.00 | 99.79 | 0.54 | 100 |

| | RaB (VdK) | | | HMR (VdK) | | | G1 (Char) | | |
|---|---|---|---|---|---|---|---|---|---|
|  | Glass | Standard deviation | Start Material | Glass | Standard deviation | Start Material | Glass | Standard deviation | Start Material |
| Na$_2$O | 3.84 | 0.17 | 3.68 | 3.84 | 0.16 | 3.55 | 0.04 | 0.04 | 0.00 |
| MgO | 24.98 | 0.20 | 25.47 | 20.69 | 0.66 | 25.70 | 24.95 | 0.60 | 25.90 |
| Al$_2$O$_3$ | 12.09 | 0.20 | 11.93 | 11.34 | 0.25 | 9.90 | 8.35 | 0.22 | 8.28 |
| SiO$_2$ | 52.29 | 0.27 | 52.70 | 52.38 | 0.47 | 51.28 | 52.98 | 0.45 | 53.60 |
| K$_2$O | 0.05 | 0.03 | | 0.05 | 0.03 | | 0.04 | 0.03 | 0.00 |
| CaO | 5.18 | 0.20 | 4.98 | 7.92 | 0.48 | 6.49 | 10.56 | 0.38 | 10.30 |
| FeO | 0.24 | 0.06 | | 1.94 | 0.16 | 1.89 | 2.17 | 0.17 | 1.44 |
| TiO$_2$ | 1.33 | 0.21 | 1.24 | 1.50 | 0.26 | 1.20 | 0.60 | 0.17 | 0.42 |
| Cr$_2$O$_3$ | 0.04 | 0.04 | | 0.03 | 0.04 | | 0.03 | 0.03 | |
| MnO | 0.02 | 0.03 | | 0.05 | 0.05 | | 0.03 | 0.03 | 0.00 |
| Total | 100.06 | | 100.00 | 99.74 | | 100.00 | 99.75 | | 99.94 |

| | G1(b) (Char) | | | HMR CaS (VdK) | | | HMC (Pep) | | |
|---|---|---|---|---|---|---|---|---|---|
|  | Glass | Standard deviation | Start Material | Glass | Standard deviation | Start Material | Glass | Standard deviation | Start Material |
| Na$_2$O | 0.04 | 0.04 | | 4.11 | 0.29 | 3.46 | 4.88 | 0.36 | 3.23 |
| MgO | 24.75 | 0.34 | 25.92 | 20.49 | 1.10 | 26.91 | 17.68 | 2.42 | 29.82 |
| Al$_2$O$_3$ | 8.71 | 0.18 | 8.28 | 10.70 | 0.39 | 9.07 | 6.32 | 0.57 | 4.53 |
| SiO$_2$ | 53.44 | 0.35 | 53. | 52.12 | 0.46 | 49.94 | 55.18 | 1.12 | 51.31 |
| K$_2$O | 0.05 | 0.03 | | 0.05 | 0.04 | | 0.05 | 0.03 | 0.00 |
| CaO | 10.97 | 0.28 | 10.3 | 9.63 | 0.41 | 7.79 | 11.59 | 0.71 | 7.38 |
| FeO | 1.44 | 0.16 | 1.44 | 1.50 | 0.14 | 1.66 | 3.71 | 0.25 | 3.12 |
| TiO$_2$ | 0.41 | 0.15 | 0.42 | 1.42 | 0.33 | 1.17 | 0.05 | 0.07 | 0.49 |
| Cr$_2$O$_3$ | 0.02 | 0.03 | | 0.03 | 0.04 | | 0.03 | 0.03 | 0.00 |
| MnO | 0.03 | 0.03 | | 0.02 | 0.04 | | 0.03 | 0.03 | 0.13 |
| Total | 99.87 | | 99.94 | 100.07 | | 100.00 | 99.51 | | 100.00 |

Table 1: Glass: glass compositions determined by electron microprobe (EMPA data for the chemistry of the glass% wt.). Start Material: nominal composition of starting materials (% wt.), EMPA Analyses: composition of melt glass (in wt %).

|  | G1 (Char) | G1(b) (Char) | ICP-HCTa (Stock) | HMR CaS (VdK) | RaB (VdK) | HAI (VdK) | PD (VdK) | HMR (VdK) | HMC (Pep) |
|---|---|---|---|---|---|---|---|---|---|
| $Na_2O$ | 0.01 | 0.01 | 0.01 | 0.01 | 0.01 | 0.18 | 0.02 | 0.02 | 0.03 |
| $MgO$ | 57.44 | 57.22 | 57.36 | 57.79 | 57.94 | 57.28 | 58.17 | 57.97 | 57.66 |
| $Al_2O_3$ | 0.05 | 0.22 | 0.13 | 0.06 | 0.84 | 0.74 | 0.04 | 0.05 | 0.03 |
| $SiO_2$ | 42.41 | 42.49 | 42.33 | 42.22 | 41.85 | 42.58 | 42.40 | 42.24 | 42.50 |
| $K_2O$ | 0.00 | 0.01 | 0.01 | 0.01 | 0.01 | 0.00 | 0.02 | 0.01 | 0.00 |
| $CaO$ | 0.30 | 0.44 | 0.15 | 0.34 | 0.18 | 0.27 | 0.25 | 0.25 | 0.37 |
| $FeO$ | 0.61 | 0.46 | 1.09 | 0.35 | 0.03 | 0.03 | 0.33 | 0.45 | 0.76 |
| $TiO_2$ | 0.02 | 0.05 | 0.03 | 0.01 | 0.08 | 0.09 | 0.00 | 0.03 | 0.04 |
| $Cr_2O_3$ | 0.00 | 0.01 | 0.01 | 0.01 | 0.01 | 0.07 | 0.02 | 0.00 | 0.00 |
| $MnO$ | 0.01 | 0.00 | 0.00 | 0.01 | 0.01 | 0.00 | 0.00 | 0.02 | 0.01 |
| Total | 100.86 | 100.91 | 101.13 | 100.79 | 100.96 | 101.25 | 101.24 | 101.06 | 101.39 |
| **Fo** | 99.41 | 99.55 | 98.94 | 99.66 | 99.97 | 99.97 | 99.69 | 99.57 | 99.27 |

Table 2: EMPA data for the minerals or (olivine) inclusion in the glass (in wt.%). Fo = forsterite content of olivine

| | | | | | | | |
|---|---|---|---|---|---|---|---|
| **NP-LMG (VdK)** | | | | | | | |
| **0-25** | 2.92 | 3.42 | 3.5 | 7.93 | 9.61 | | 11.81 |
| **25-63** | 2.85 | 3.42 | 3.5 | 7.93 | 9.55 | | |
| **63-125** | 2.82 | 3.41 | 3.5 | 7.94 | 9.54 | | |
| **125-250** | 2.8 | 3.41 | 3.5 | 7.92 | 9.56 | | |
| **NC (Pep)** | | | | | | | |
| **0-25** | 2.94 | 3.42 | 3.5 | 7.98 | 9.74 | | 11.87 |
| **25-63** | 2.93 | 3.41 | 3.5 | 7.95 | 9.65 | | |
| **63-125** | 2.83 | 3.42 | 3.5 | 7.96 | 9.67 | | |
| **125-250** | 2.82 | 3.41 | 3.5 | 7.91 | 9.66 | | |
| **CBC (Pep)** | | | | | | | |
| **0-25** | 2.94 | 3.42 | 3.5 | 7.98 | 9.82 | | 11.83 |
| **25-63** | 2.87 | 3.42 | 3.5 | 7.95 | 9.76 | | |
| **63-125** | 2.82 | 3.42 | 3.5 | 7.95 | 9.75 | | |
| **125-250** | 2.82 | 3.42 | 3.5 | 8 | 9.77 | | |
| **CB (VdK)** | | | | | | | |
| **0-25** | 2.95 | 3.42 | 3.5 | 8 | 9.73 | | 11.87 |
| **25-63** | 2.87 | 3.42 | 3.5 | 7.97 | 9.77 | | |
| **63-125** | 2.84 | 3.42 | 3.5 | 7.98 | 9.76 | | |
| **125-250** | 2.82 | 3.41 | 3.49 | 7.98 | 9.78 | | |
| **NVPa(Stock)** | | | | | | | |
| **0-25** | 2.92 | 3.41 | 3.50 | 7.96 | | 10.74 | 11.74 |
| **25-63** | 2.86 | 3.42 | 3.50 | 7.95 | 9.52 | 10.54 | |
| **63-125** | 2.82 | 3.41 | 3.50 | 7.91 | 9.66 | 10.50 | |
| **125-250** | 2.81 | 3.41 | 3.50 | 7.93 | 9.60 | 10.50 | |
| **NP-HMG (VdK)** | | | | | | | |
| **0-25** | 2.95 | 3.42 | 3.5 | 8.07 | 9.74 | | 11.95 |
| **25-63** | 2.9 | 3.41 | 3.5 | 8.04 | 9.77 | | |
| **63-125** | 2.85 | 3.41 | 3.5 | 8.05 | 9.77 | | |
| **125-250** | 2.84 | 3.42 | 3.5 | 8.08 | 9.77 | | |
| **G2 (Char)** | | | | | | | |
| **0-25** | 2.95 | 3.41 | 3.50 | 8.07 | 10.57 | | 11.92 |
| **25-63** | 2.90 | 3.41 | 3.50 | 8.05 | 10.21 | | |
| **63-125** | 2.84 | 3.41 | 3.50 | 8.04 | 10.28 | | |
| **125-250** | 2.82 | 3.40 | 3.50 | 8.03 | 10.29 | | |

Table 3: Band position of major features in the diffuse reflectance spectra of all four sieved size fractions (in μm).

| | | | | | | | |
|---|---|---|---|---|---|---|---|
| **HAI (VdK)** | | | | | | | |
| **0-25** | 2.94 | 3.42 | 3.5 | 8.21 | 9.95 | | 11.94 |
| **25-63** | 2.89 | 3.42 | 3.5 | 8.15 | 9.92 | | |
| **63-125** | 2.84 | 3.42 | 3.5 | 8.15 | 9.92 | | |
| **125-250** | 2.82 | 3.42 | 3.5 | 8.14 | 9.92 | | |
| **IC (Pep)** | | | | | | | |
| **0-25** | 2.92 | 3.42 | 3.5 | 8.13 | 9.91 | | 11.99 |
| **25-63** | 2.86 | 3.42 | 3.5 | 8.13 | 9.9 | | |
| **63-125** | 2.84 | 3.41 | 3.5 | 8.14 | 9.91 | | |
| **125-250** | 2.82 | 3.41 | 3.5 | 8.16 | 9.94 | | |
| **IT (VdK)** | | | | | | | |
| **0-25** | 2.93 | 3.42 | 3.5 | 8.19 | 9.94 | | 12.06 |
| **25-63** | 2.85 | 3.42 | 3.5 | 8.2 | 9.96 | | |
| **63-125** | 2.84 | 3.42 | 3.5 | 8.16 | 9.94 | | |
| **125-250** | 2.82 | 3.42 | 3.5 | 8.16 | 9.93 | | |
| **PD (VdK)** | | | | | | | |
| **0-25** | 2.95 | 3.42 | 3.5 | 8.26 | 10 | | 12.07 |
| **25-63** | 2.89 | 3.42 | 3.5 | 8.23 | 9.88 | | |
| **63-125** | 2.84 | 3.42 | 3.5 | 8.21 | 9.92 | | |
| **125-250** | 2.84 | 3.41 | 3.5 | 8.2 | 9.92 | | |
| **ICP-(HCTa)** | | | | | | | |
| **0-25** | 2.92 | 3.42 | 3.50 | 8.15 | 10.66 | | 11.95 |
| **25-63** | 2.84 | 3.42 | 3.50 | 8.12 | 10.30 | | |
| **63-125** | 2.82 | 3.42 | 3.50 | 8.12 | 10.24 | | |
| **125-250** | 2.82 | 3.41 | 3.50 | 8.12 | 10.31 | | |
| **RaB (VdK)** | | | | | | | |
| **0-25** | 3.00 | 3.42 | 3.50 | 8.23 | 9.91 | | 12.16 |
| **25-63** | 2.87 | 3.42 | 3.50 | 8.24 | 9.91 | | |
| **63-125** | 2.83 | 3.42 | 3.50 | 8.21 | 9.92 | | |
| **125-250** | 2.82 | 3.41 | 3.50 | 8.21 | 9.94 | | |

Table 3 ff

| | | | | | | | | | | | |
|---|---|---|---|---|---|---|---|---|---|---|---|
| **HMR (VdK)** | | | | | | | | | | | |
| **0-25** | 2.95 | 3.42 | 3.5 | 8.32 | | 10.12 | 10.68 | | 12.2 | | |
| **25-63** | 2.9 | 3.42 | 3.5 | 8.26 | 9.9 | 10.12 | 10.49 | 11.9 | | 16.25 | 16.51 |
| **63-125** | 2.84 | 3.42 | 3.5 | 8.22 | | 10.12 | 10.45 | 11.9 | | | 16.44 |
| **125-250** | 2.84 | 3.41 | 3.5 | 8.2 | 9.9 | 10.12 | 10.47 | 11.9 | | 16.34 | 16.66 |
| **G1 (Char)** | | | | | | | | | | | |
| **0-25** | 2.93 | 3.42 | 3.5 | 8.21 | 9.86 | 10.13 | 10.59 | 10.68 | 12.15 | | |
| **25-63** | 2.84 | 3.42 | 3.5 | 8.21 | 9.88 | 10.12 | 10.5 | | | | |
| **63-125** | 2.82 | 3.42 | 3.5 | 8.2 | | 10.13 | 10.49 | | | | |
| **125-250** | 2.82 | 3.42 | 3.5 | 8.22 | | 10.12 | 10.46 | | | | |
| **HMR-CaS (VdK)** | | | | | | | | | | | |
| **0-25** | 2.95 | 3.41 | 3.5 | 8.4 | | 10.12 | 10.67 | | 12.23 | 16.15 | |
| **25-63** | 2.9 | 3.41 | 3.5 | 8.31 | 9.87 | 10.12 | 10.51 | 11.9 | | 16.15 | 16.44 |
| **63-125** | 2.85 | 3.41 | 3.5 | 8.26 | 9.85 | 10.12 | 10.49 | 11.9 | | 16.26 | 16.48 |
| **125-250** | 2.84 | 3.41 | 3.5 | 8.27 | | 10.13 | 10.5 | 11.9 | | 16.36 | |
| **HMC (Pep)** | | | | | | | | | | | |
| **0-25** | 2.94 | 3.42 | 3.5 | 8.38 | 9.8 | 10.11 | | 11.21 | 12.25 | 16.22 | |
| **25-63** | 2.84 | 3.42 | 3.5 | 8.31 | | 10.13 | 10.56 | 11.9 | | 16.15 | |
| **63-125** | 2.82 | 3.42 | 3.5 | 8.27 | | 10.13 | 10.6 | 11.89 | | 16.28 | |
| **125-250** | 2.82 | 3.42 | 3.5 | 8.26 | | 10.12 | 10.5 | 11.86 | | 16.3 | |

Table 3 ff

**Left column:**

| Sample / Spot | B1 | B2 | B3 |
|---|---|---|---|
| **NP-LMG (VdK)** | | | |
| 1 | 7.9 | 9.62 | |
| 2 | 7.88 | 9.55 | |
| **NC (Pep)** | | | |
| 1 | 7.94 | 9.73 | |
| **CBC (Pep)** | | | |
| 1 | 7.94 | 9.89 | |
| 2 | 7.96 | 9.75 | |
| **CB (VdK)** | | | |
| 1 | 8.01 | 9.86 | |
| 2 | 7.99 | 9.98 | |
| **NVPa (Stock)** | | | |
| 1 | 7.92 | | 10.58 |
| 2 | 7.92 | | 10.52 |
| **N- HMG (VdK)** | | | |
| 1 | 8.04 | 9.82 | |
| 2 | 8.03 | 9.84 | |
| **G2 (Char)** | | | |
| 1 | 8.07 | | 10.4 |
| **HAI (VdK)** | | | |
| 1 | 8.18 | 10.02 | |
| 2 | 8.13 | 9.96 | |
| **IC (Pep)** | | | |
| 1 | 8.14 | 9.94 | |
| 2 | 8.11 | 9.95 | |
| **IT (VdK)** | | | |
| 1 | 8.13 | 9.99 | |
| 2 | 8.18 | 10.12 | |
| **PD (VdK)** | | | |
| 1 | 8.21 | 9.96 | 10.06 |
| 2 | 8.21 | 9.98 | |

**Right column:**

| Sample / Spot | B1 | B2 | B3 | B4 | B5 | B6 |
|---|---|---|---|---|---|---|
| **ICP-HCTa (Stock)** | | | | | | |
| 1 | 8.11 | 9.51 | 9.91 | 10.43 | 10.81 | |
| **RaB (VdK)** | | | | | | |
| 1 glass? | 8.15 | 9.99 | | | | |
| 2 glass? | 8.11 | 9.98 | | | | |
| 3 | 8.12 | 9.95 | 10.14 | | | |
| 4 | 8.17 | 9.94 | 10.13 | 10.44 | | |
| **HMR (VdK)** | | | | | | |
| 1 | 8.17 | 9.63 | 9.77 | 10.16 | | 10.82 |
| 2 glass ? | 8.15 | 9.98 | | 10.16 | | |
| 3 | 8.2 | | | 10.15 | | 10.7 |
| 4 | 8.15 | | | 10.16 | 10.47 | |
| **G1 (Char)** | | | | | | |
| 1 glass ? | 8.22 | | 10.11 | | | |
| 2 glass ? | 8.25 | | 10.1 | | | |
| 3 | 8.18 | | 10.16 | 10.81 | 10.98 | |
| 4 | 8.19 | | 10.16 | | 10.81 | |
| 1(b) glass ? | 8.2 | | 10.12 | | | |
| **HMR-CaS (VdK)** | | | | | | |
| 1 | 8.25 | | 10.16 | 10.71 | | |
| 2 glass ? | 8.23 | | 10.13 | | | |
| 3 | 8.28 | | 10.15 | 10.77 | 10.87 | |
| **HMC (Pep)** | | | | | | |
| 1 | 8.22 | 9.63 | 10.15 | 10.89 | 11.9 | |
| 2 | 8.2 | | 10.15 | 11 | 11.9 | |

Table 4: Band positions of selected spots analyzed in-situ with micro-FTIR (aperture 256×256μm) on polished sections of the samples

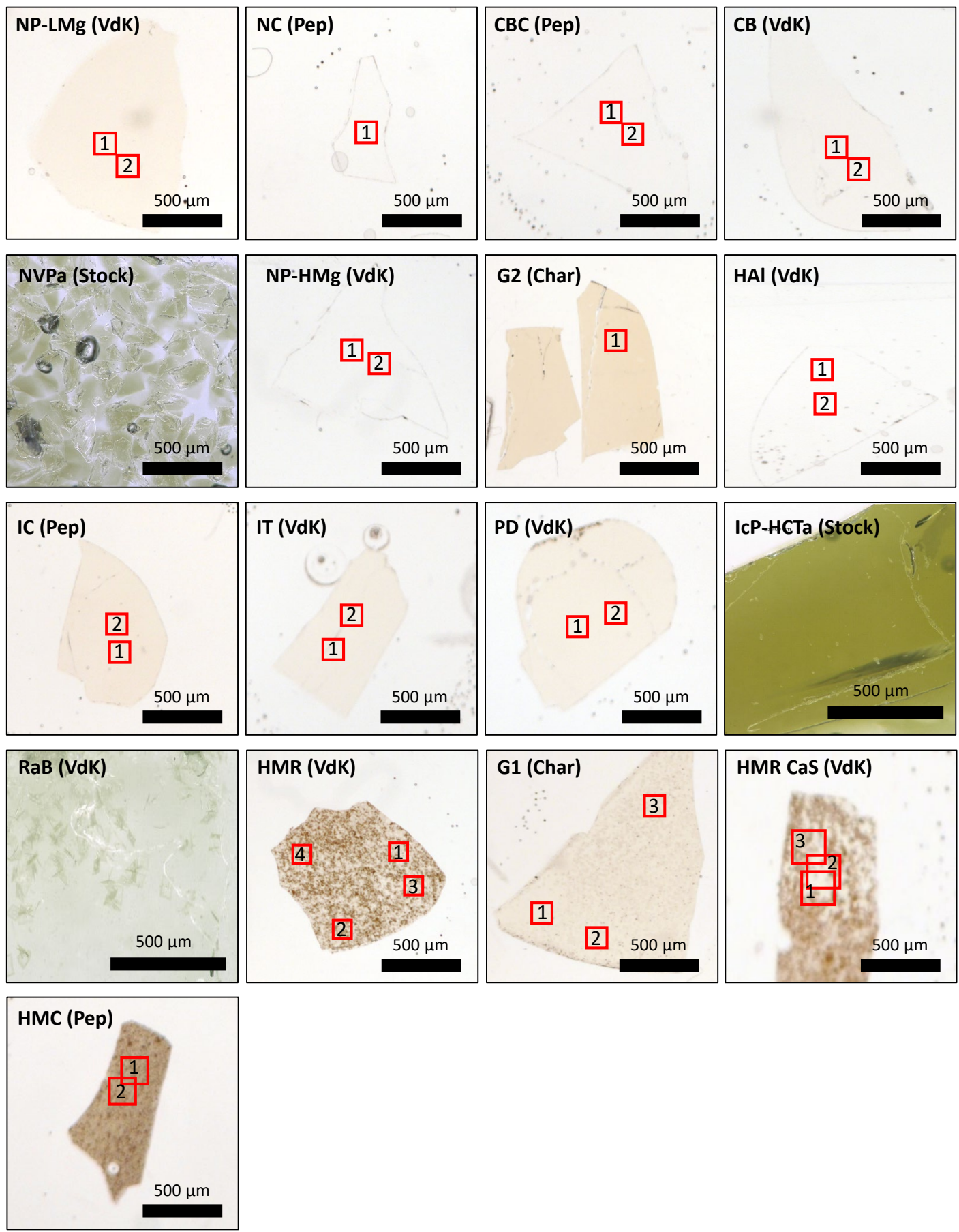

Fig. 1

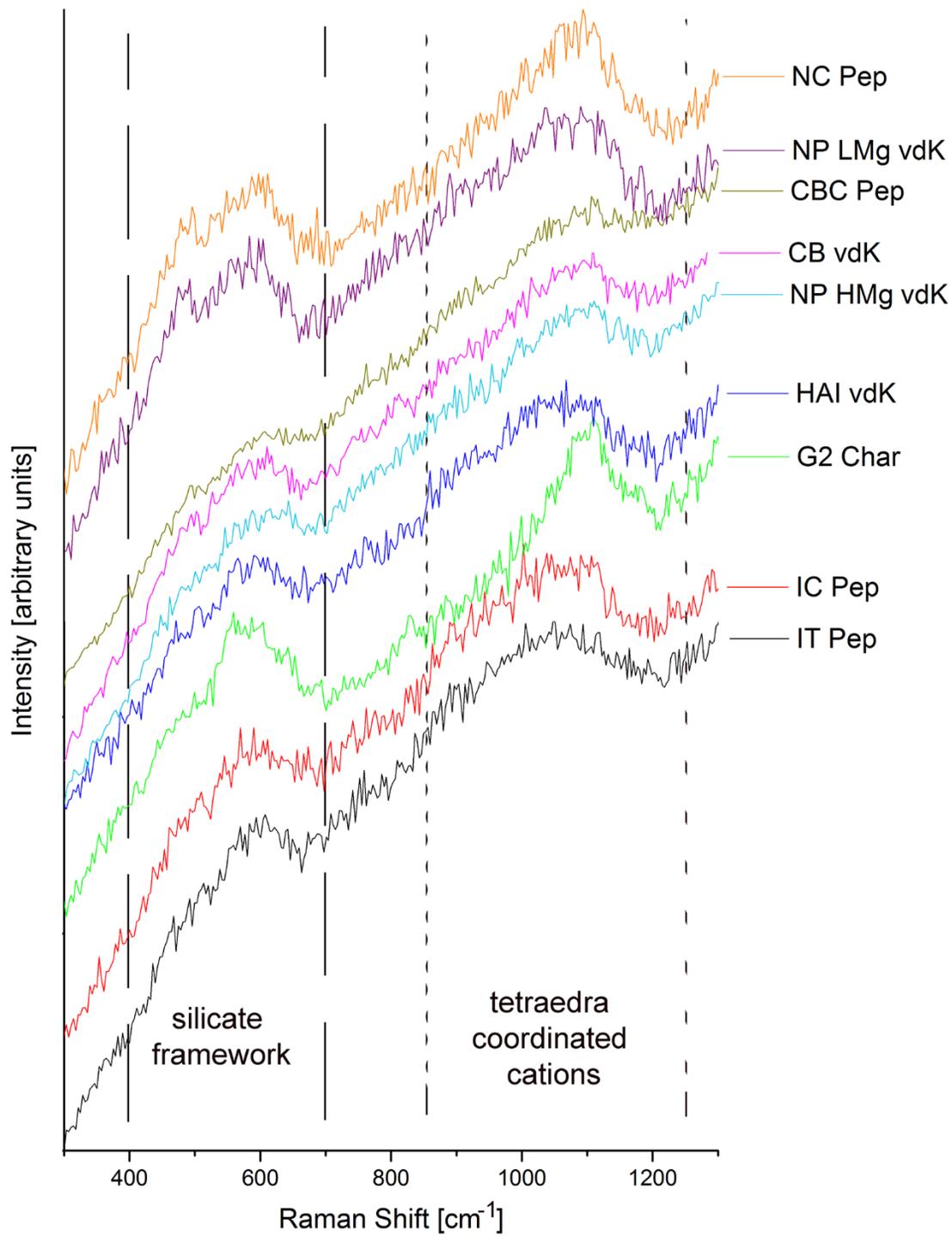

Fig. 2a

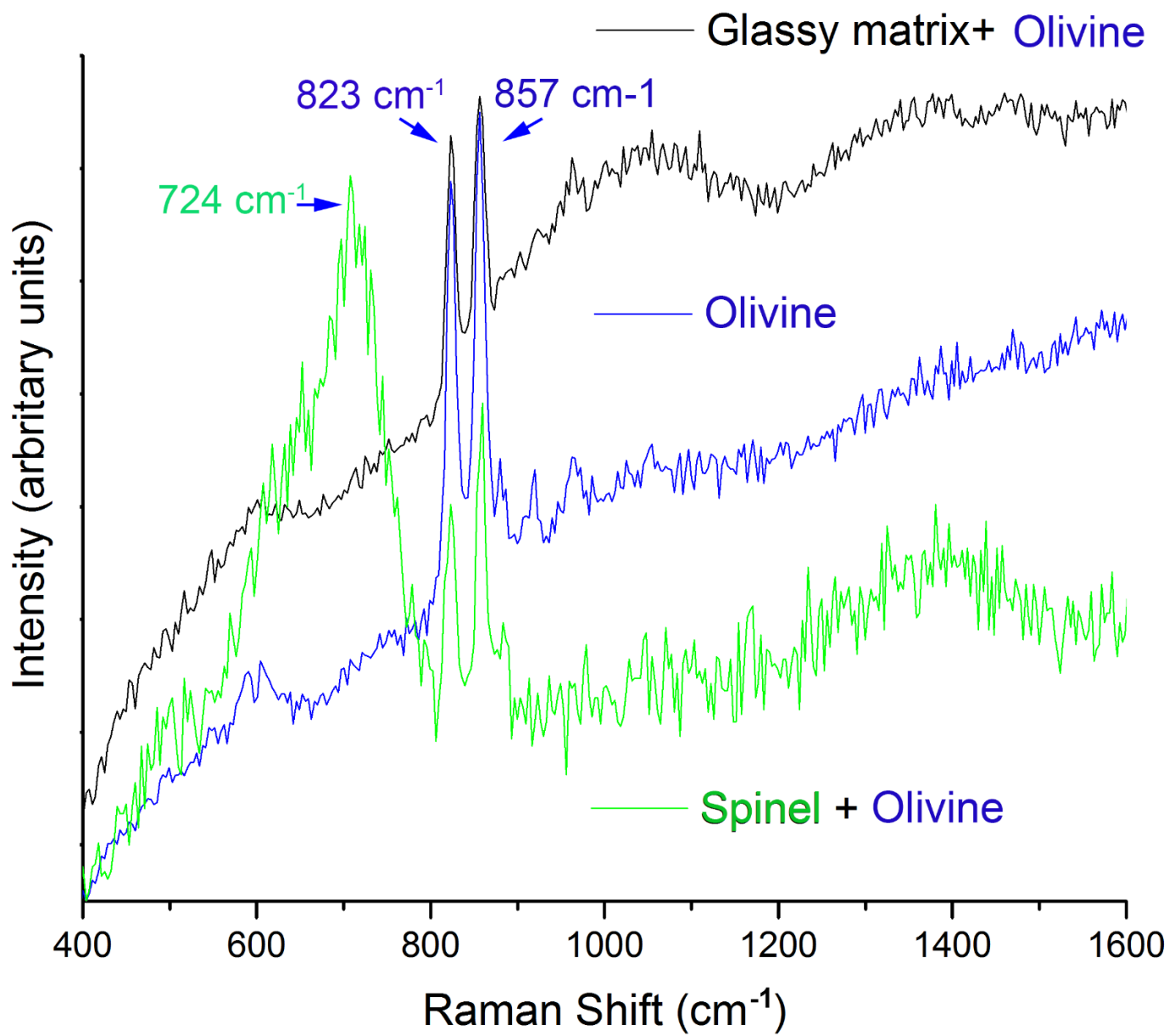

Fig. 2b

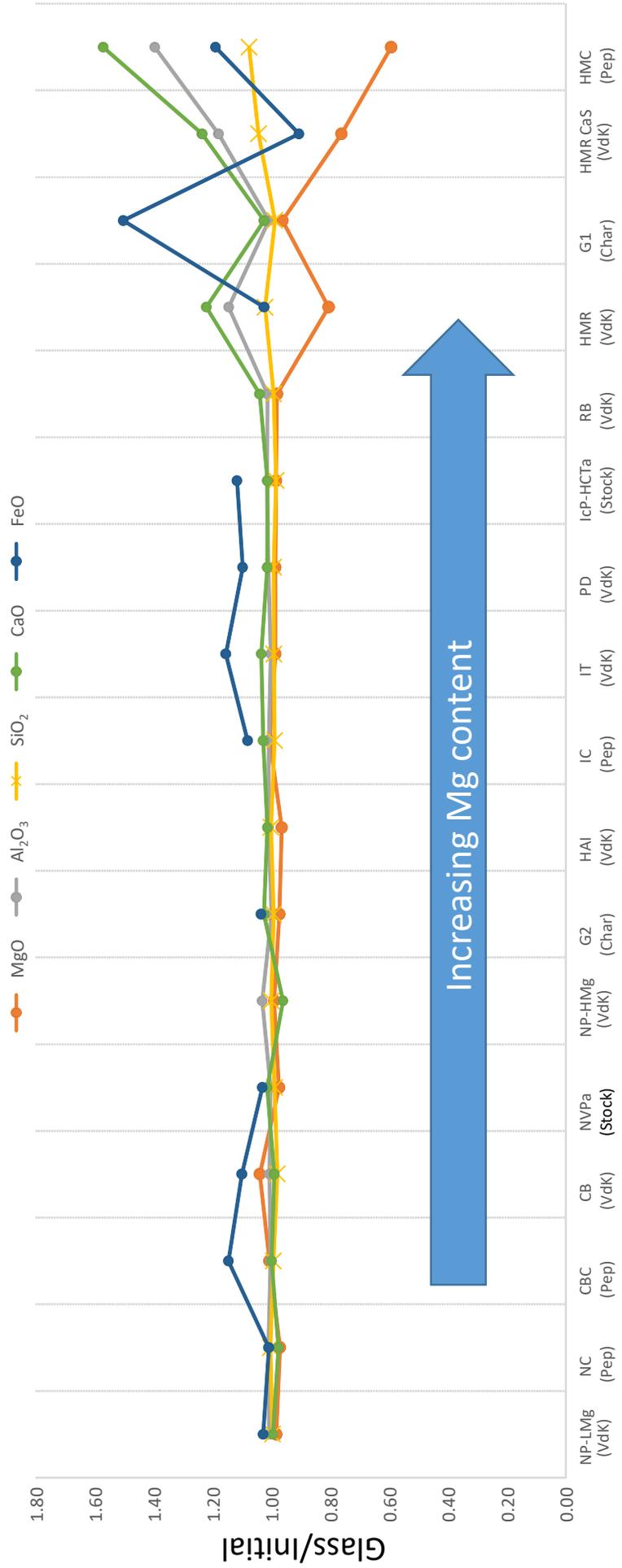

Fig. 3

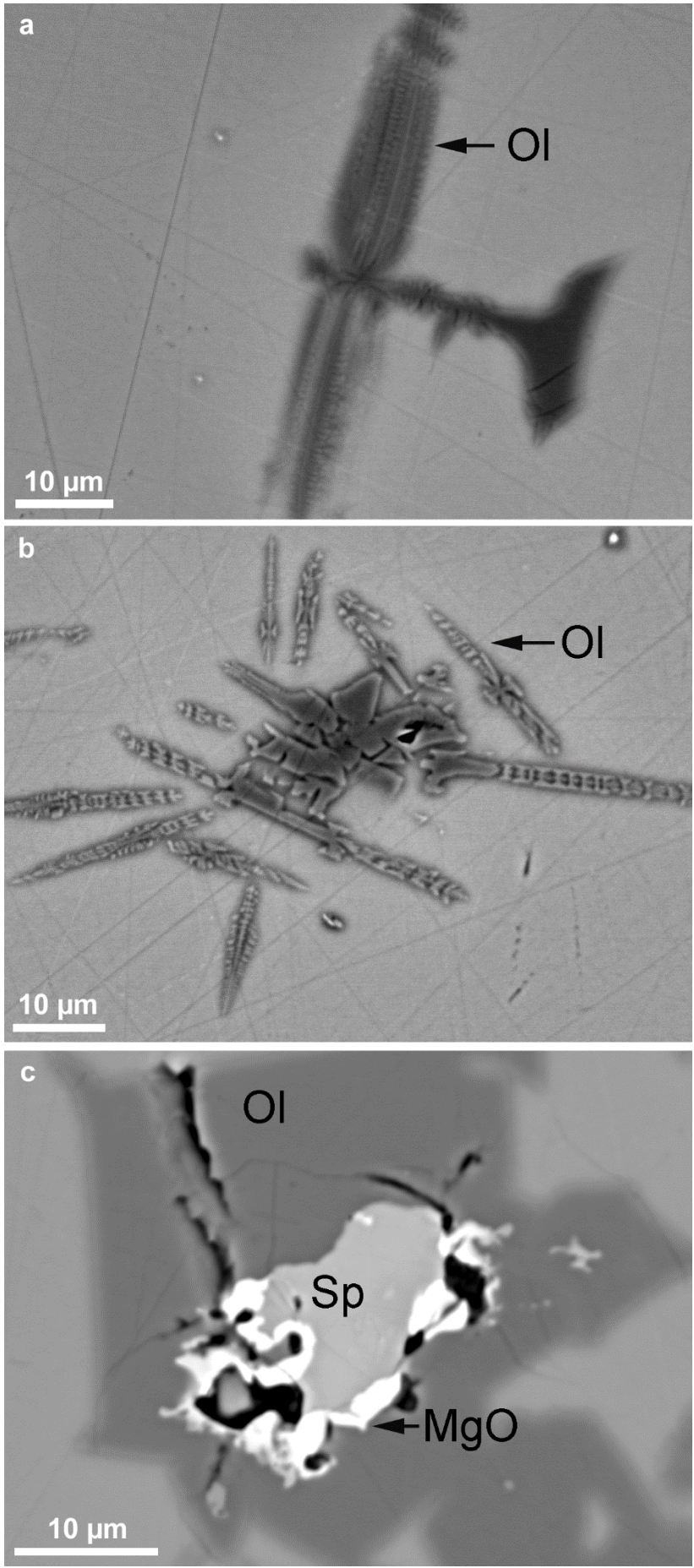

Fig. 4

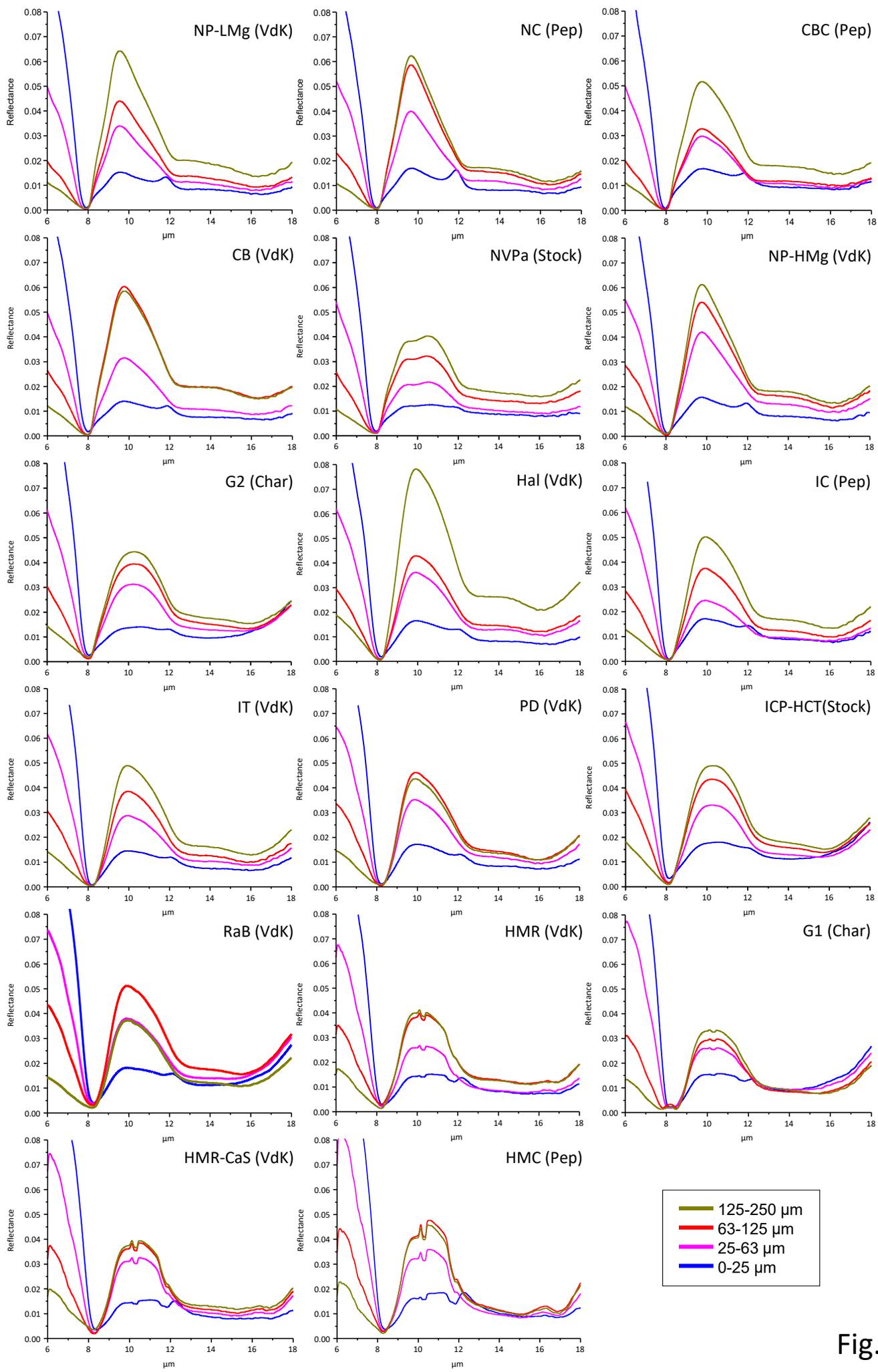

Fig. 5

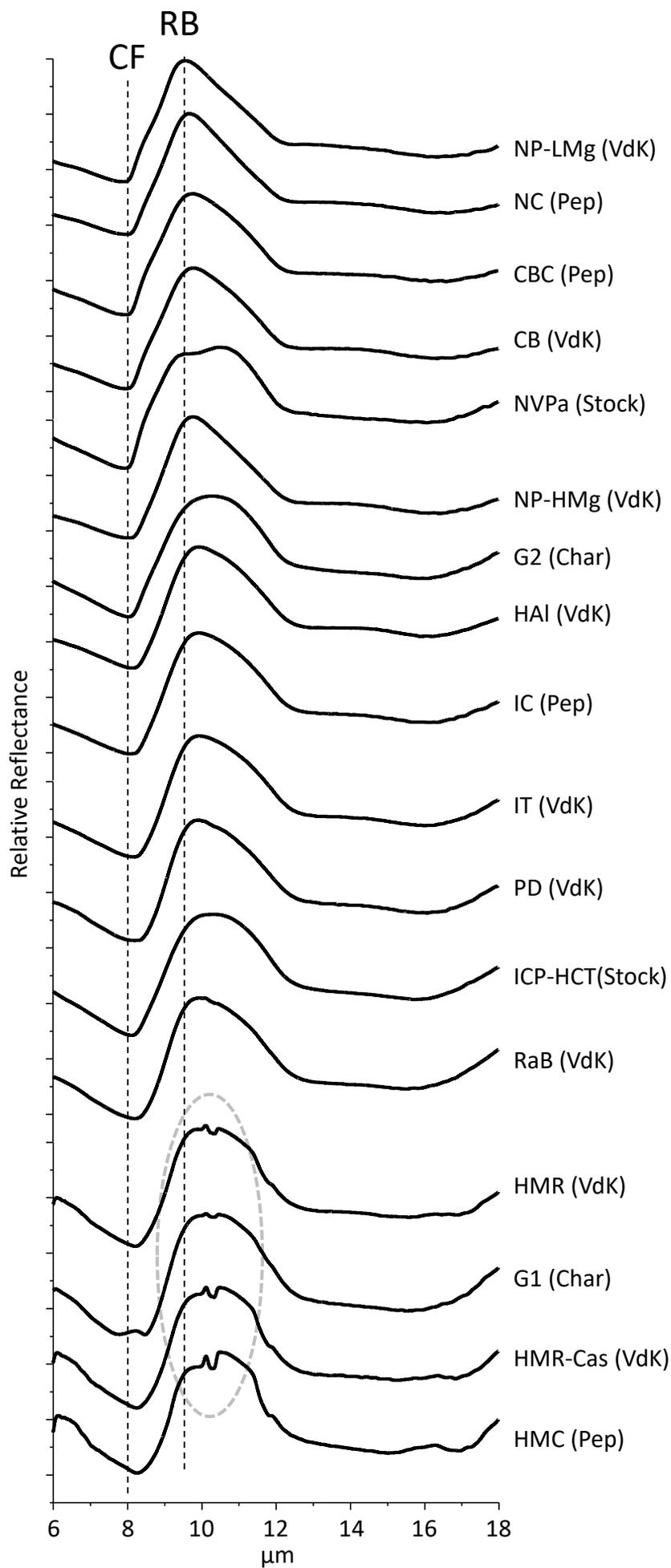

Fig. 6a

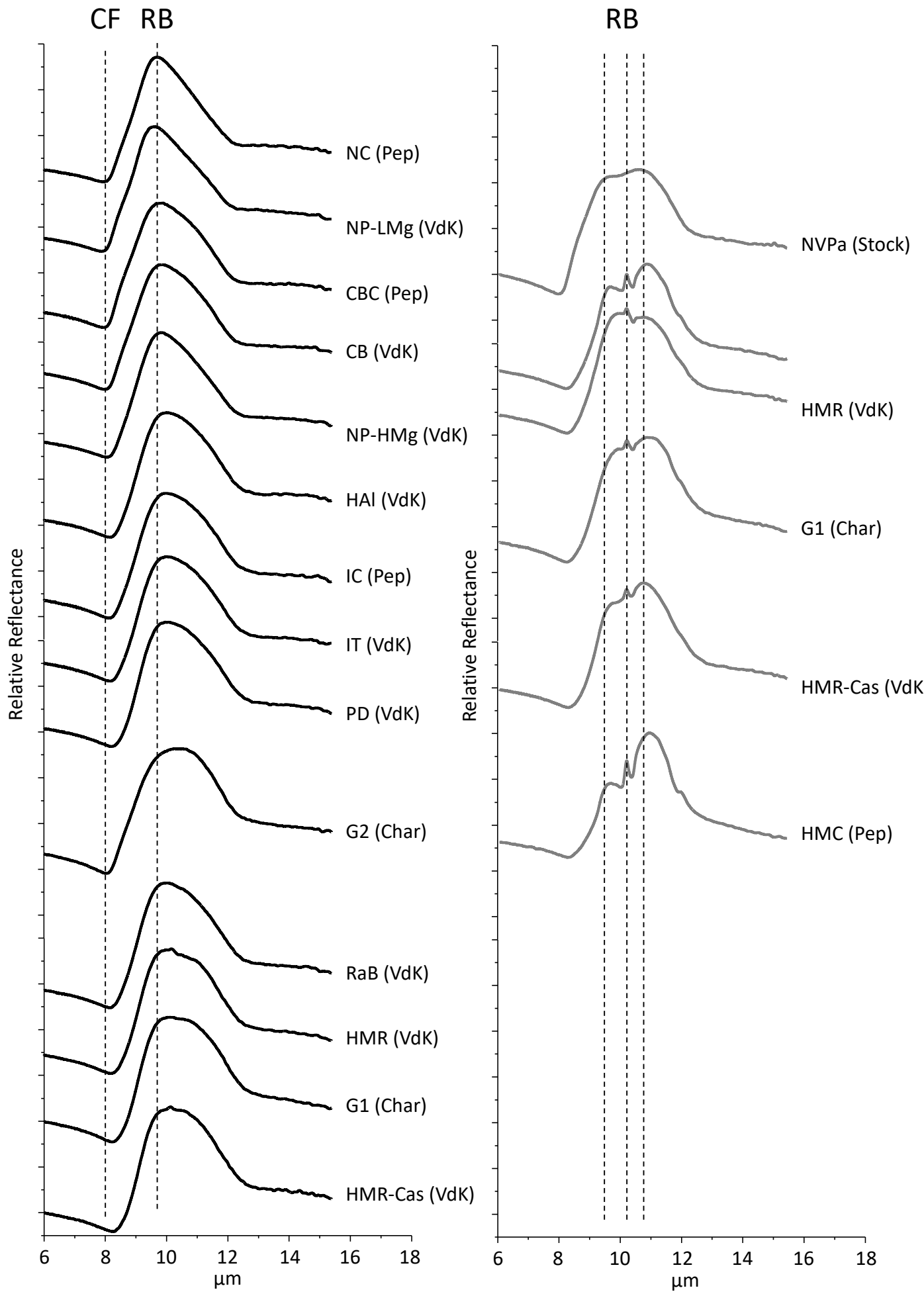

Fig. 6b

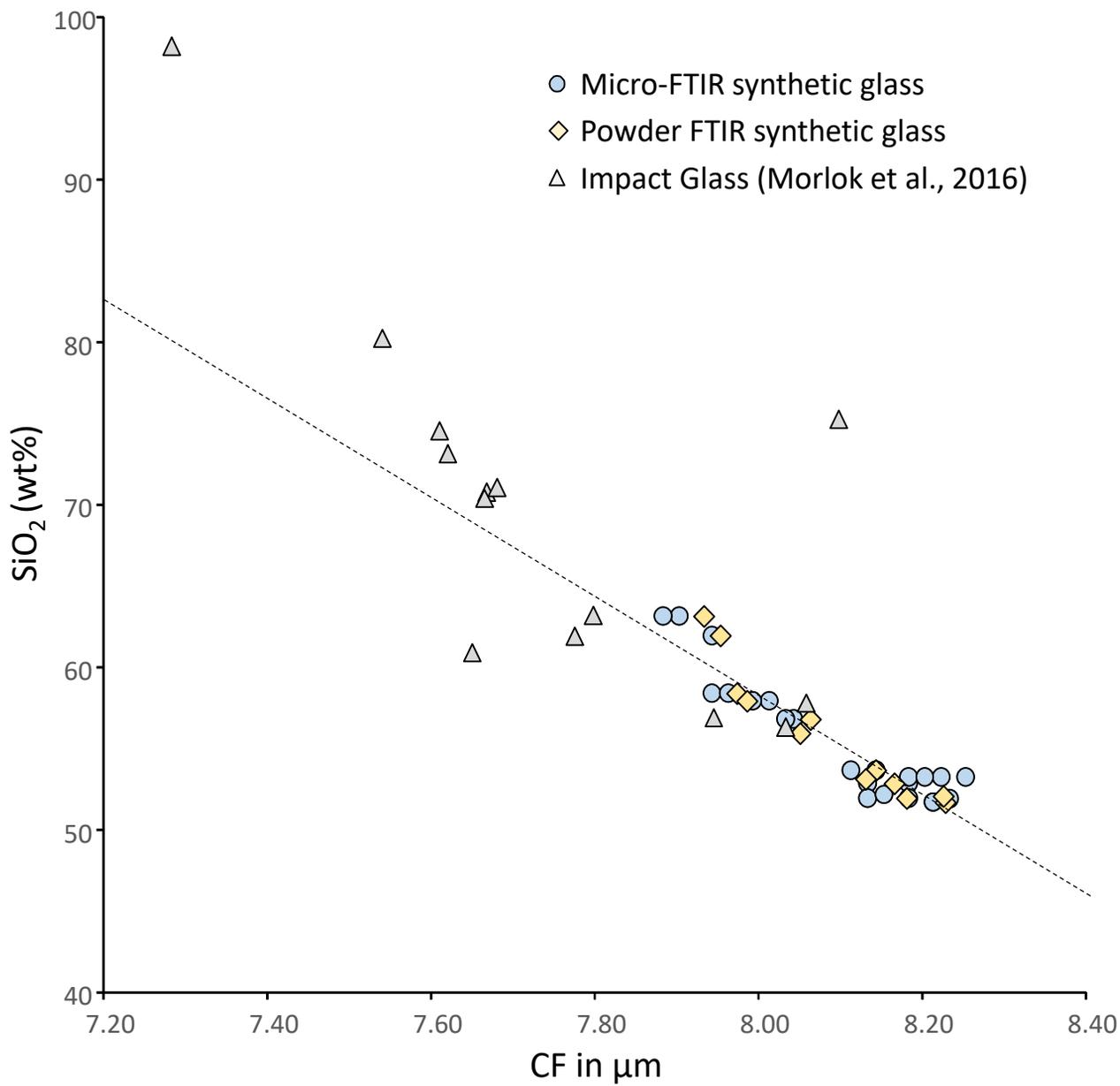

Fig. 7

Fig. 8

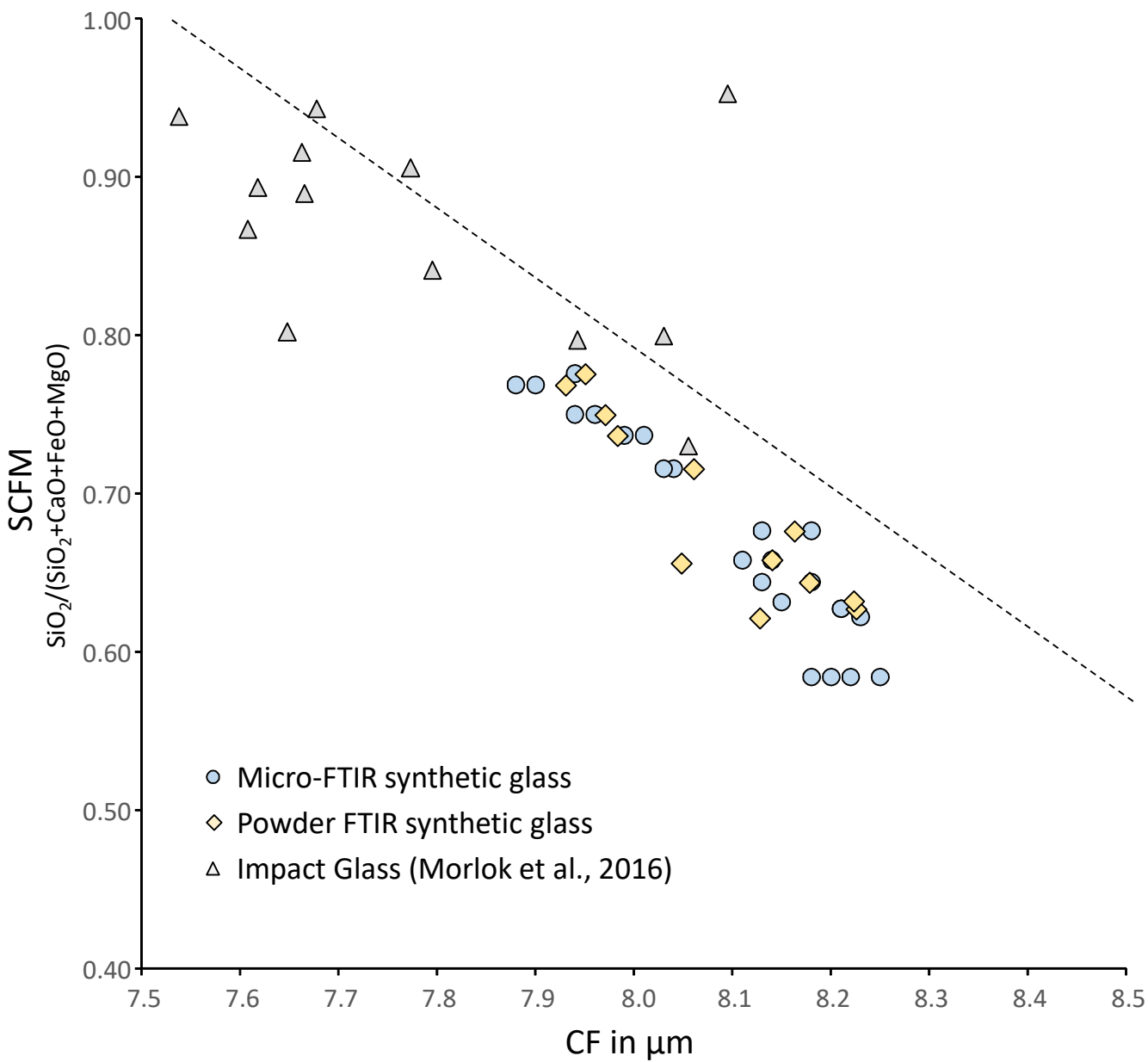

Fig. 9

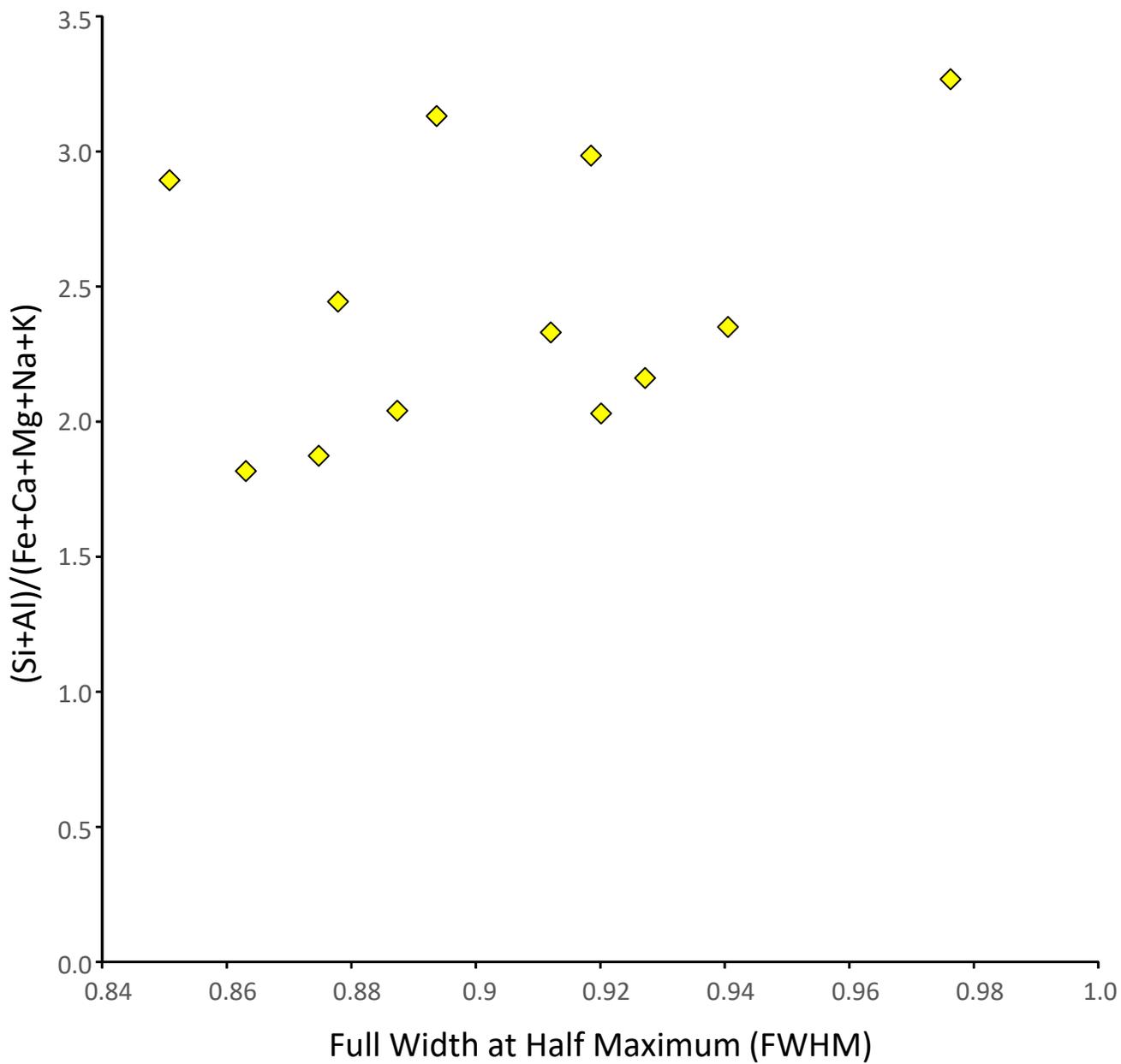

Fig. 10

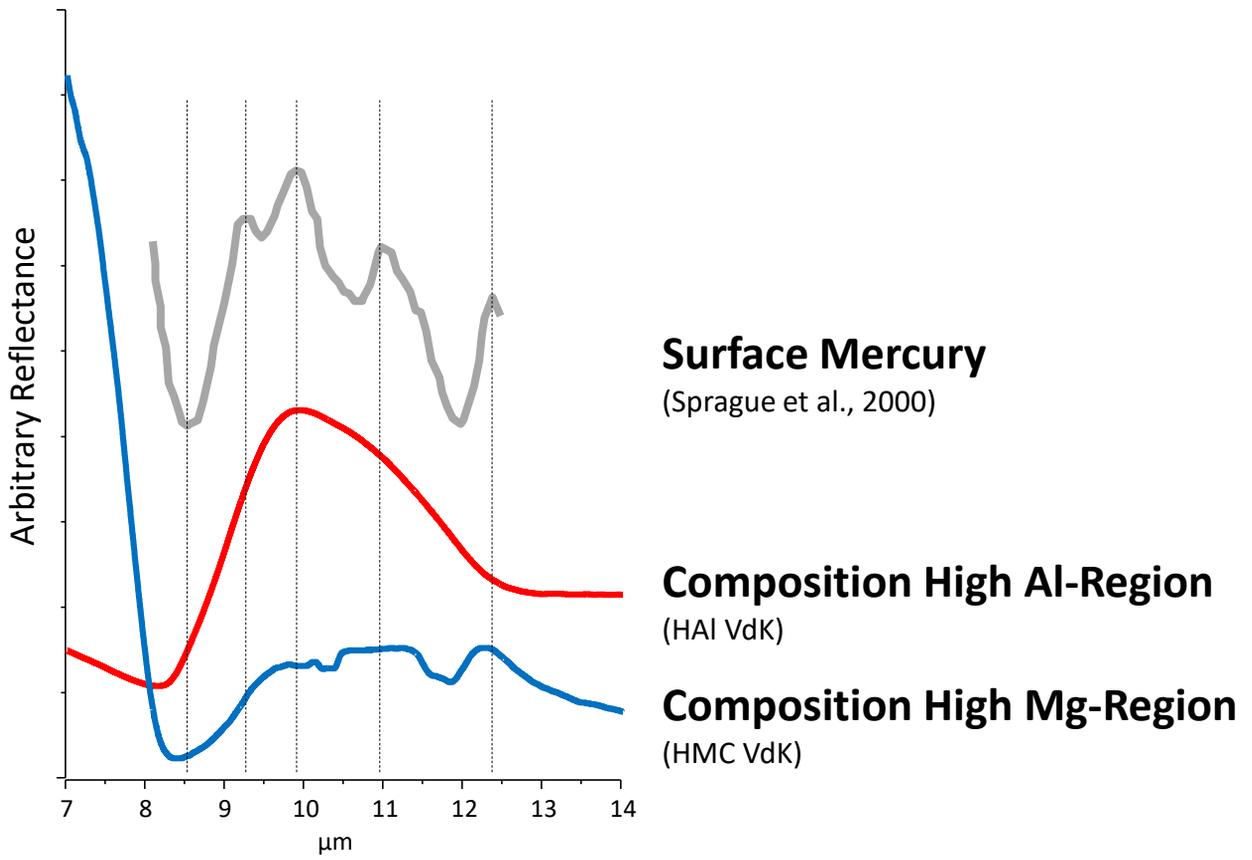

**Surface Mercury**
(Sprague et al., 2000)

**Composition High Al-Region**
(HAl VdK)

**Composition High Mg-Region**
(HMC VdK)

Fig. 11